\documentclass[12pt]{article}
\usepackage{amsmath, amssymb, slashed, epsf, color, graphicx,cite}
\usepackage{epsfig}

\def\dslash{\partial \hspace*{-0.5em}/\hspace*{0.3em}}

\def\dslash{\partial \hspace*{-0.5em}/\hspace*{0.3em}}
\def\met{E_T \hspace*{-1.1em}/\hspace*{0.5em}}

\def\fb{{\rm ~fb}^{-1}}
\def\mchi{m_\chi}
\def\gev{{\rm ~GeV}}
\def\sv{\langle \sigma v \rangle_{\rm F.O.}}

\newcommand{\sigsip}{\ensuremath{\sigma^{\rm{SI}}_p}}

\newcommand{\sigsdn}{\ensuremath{\sigma^{\rm{SD}}_n}}
\newcommand{\mev}{\ensuremath{\,\mathrm{MeV}}}
\newcommand{\tev}{\ensuremath{\,\mathrm{TeV}}}

\begin{document}

\begin{titlepage}
\begin{center}

\hfill IPMU14-0155 \\

\vspace{1.5cm}
{\large\bf Singlet Majorana fermion dark matter:\\
a comprehensive analysis in effective field theory}

\vspace{1.5cm}
{\bf Shigeki Matsumoto\footnote{email: shigeki.matsumoto@ipmu.jp}, }
{\bf Satyanarayan Mukhopadhyay\footnote{email: satya.mukho@ipmu.jp}}\\
{\bf and}
{\bf Yue-Lin Sming Tsai\footnote{email: yue-lin.tsai@ipmu.jp}}\\

\vspace{1.0cm}
{\it
{Kavli IPMU (WPI), The University of Tokyo, Kashiwa, Chiba 277-8583, Japan}} \\[2cm]

\abstract{We explore a singlet Majorana fermion dark matter candidate using an effective field theory (EFT) framework, respecting the relations imposed by the standard model $SU(3)_C \times SU(2)_L \times U(1)_Y$ gauge invariance among different couplings. All operators of dimension-5 and dimension-6, forming a complete basis, are taken into account at the same time, keeping in view ultraviolet completions which can give rise to more than one operator at a time. If in addition CP-conservation is assumed, the remaining parameter space, where an EFT description is valid, is found to be rather restricted after imposing constraints from relic abundance, direct, indirect and collider searches. On including the CP-violating dimension-5 operator, $(\overline{\chi}i \gamma_5 \chi) (H^\dagger H)$, a significantly larger parameter space opens up. We use the profile likelihood method to map out the remaining landscape of such a DM scenario. The reach of future searches using ton-scale direct detection experiments, an $e^+ e^-$ collider like the proposed ILC and limits from future gamma-ray observations are also estimated.}

\end{center}
\end{titlepage}
\setcounter{footnote}{0}

\section{Introduction}
\label{sec1}
The evidence for non-baryonic dark matter (DM), which makes up more than $80\%$ of the matter content in the Universe, is still entirely from its gravitational interactions and there is no convincing evidence so far of any couplings of the DM particle with the standard model (SM) sector~\cite{trimble,review}. Dedicated efforts for more than a decade in underground DM direct detection experiments as well as the search for DM pair-annihilation products in gamma rays, antiparticles and neutrinos have now considerably restricted the possible strength of such couplings. In fact, the dual requirements of obtaining a thermal relic abundance as required by the WMAP and Planck measurements, and a low enough spin-independent (SI) scattering rate with nuclei to be consistent with the impressive bounds from experiments like XENON100 and LUX, have cornered several possible models for particle DM. 

From the theoretical viewpoint there are two possible roadmaps to explore weakly interacting massive particles (WIMP). One is to adopt a beyond SM (BSM) scenario which is motivated by some other particle physics considerations (for e.g.,  the naturalness problem of the electroweak scale) and can furnish a viable DM candidate. The most well-known example in this class is the lightest neutralino in the R-parity conserving minimal supersymmetric extension of the SM. The alternative possibility is to study DM interactions with the SM sector in an effective field theory (EFT) set-up~\footnote{In this study, an EFT is described by all the renormalizable and a complete set of higher-dimensional operators (upto a given dimension) respecting the symmetries of the SM, with arbitrary low-energy co-efficients. This does not include matching to ultraviolet complete theories or running of the operator coefficients as in the original considerations of EFT's~\cite{Georgi}.}. Needless to say, these two approaches are not completely independent.  If only the DM particle is much lighter than the other new states in the model, at energies  below the mass scale of the heavier states, any such BSM scenario is amenable to an EFT description.  In this paper, we shall focus on the latter approach. To begin our discussion of an EFT for the DM particle, we first need to fix its spin and its quantum numbers under the SM gauge group of $SU(3)_{\rm C} \times SU(2)_{\rm L} \times U(1)_{\rm Y}$. We take the DM to be a spin-$1/2$ Majorana fermion, and a singlet under the SM gauge interactions.  

The low-energy EFT of Majorana fermion DM has been studied on several occasions in the context of direct, indirect and collider experiments~\cite{Hooper,maverick,shigeki-night,Ibe,Pires,Kopp-photon,Cao,Shoemaker,Kingman,Kopp-LHC,Biplob,Wang,Kolb-gauge,Yaguna,Luca,Kolb-CPC-CPV,Sacha-Z,Man,validity-dolan,validity-riotto,Csaba}. The purpose of this paper is to improve upon the previous studies, update them in the light of recent data in all frontiers, and to perform a complete profile likelihood analysis of the EFT to determine its currently allowed parameter space. The directions in which our study goes beyond the previous approaches to the problem are the following:

\begin{enumerate}
\item We consider {\em all} operators of dimension-5 and dimension-6 allowed by the symmetries of the theory {\em at the same time} in our analysis. Not only is this justified in the spirit of an EFT, generically UV completions can lead  to the presence of more than one operator at a time in the low energy effective theory. Therefore, we work with a complete basis of all dimension-5 and dimension-6 operators. As is well-known, often there are redundancies among different higher dimensional operators which are related by the equations of motion (EOM). We have enumerated such operators as well, and have demonstrated how they can be eliminated with the help of the EOM.

\item A consequence of the above consideration is that given the DM mass, the relic density ($\Omega h^2$) requirement does not fix the co-efficient of an operator, neither does the $\Omega h^2 <0.1 $ condition put a lower bound on the same. All operators contribute additively to the annihilation cross-section at freeze-out, modulo interference effects which can be negative. These interference effects are entirely missed if we only consider one operator at a time. 

\item We write down the effective Lagrangian by respecting the SM gauge invariance under $SU(3)_{\rm C} \times SU(2)_{\rm L} \times U(1)_{\rm Y}$. As a result, the relations between different DM couplings, as implied by the SM gauge invariance are taken into account. 

\item  A complete likelihood analysis is performed by including the requirements of relic density (from Planck~\cite{Planck}), direct detection (the recent LUX~\cite{LUX} and XENON100~\cite{XENON} results), indirect detection (Fermi-LAT dwarf spheroidal observations~\cite{Fermi} and IceCube~\cite{Ice} limits) and collider (Z-boson invisible width~\cite{LEPZwidth} and monophoton search limits from LEP~\cite{LEP-monophoton}  and Higgs invisible width and monojet+ missing energy search limits from LHC~\cite{CMS}). We also properly take into account the uncertainties coming from each of these measurements in our likelihood. 
\end{enumerate}

The remaining sections of the paper are organized as follows. In section 2, we describe the basis of dimension-5 and 6 operators considered by us, electroweak symmetry breaking (EWSB) effects and the range of validity of the EFT calculations. The experimental constraints from direct, indirect and collider searches are detailed in section 3, which also contains specifics about the construction of the likelihood function including the treatment of uncertainties. Section 4 is devoted to the role of individual operators in determining the thermal component of the DM relic abundance. Our results in the CP-conserving scenario are discussed in section 5, followed by the prospects of future experiments. The case including the CP-violating dimension-5 operator is discussed in section 6. Section 7 summarizes our findings. Finally, the construction of the likelihood function for direct detection experiments, and the validation of the LHC monojet simulation framework can be found in Appendices A and B respectively.

\section{The effective field theory framework}
\label{sec2}
In this section, we describe our EFT framework, in which the low-energy degrees of freedom consist of the SM particles and the Majorana fermion DM field $\chi$. All interactions of the DM field with the SM sector are encoded by higher-dimensional operators, with a suppression scale $\Lambda$, which is the mass scale of the heavy fields integrated out to obtain the following low-energy Lagrangian:
\begin{equation}
\mathcal{L} = \mathcal{L}_{\rm SM} + \frac{1}{2} \overline{\chi}\left(i\dslash-M_{\chi} \right)\chi+\mathcal{L}_5+\mathcal{L}_6+\mathcal{L}_{\geq 7},
\end{equation}
where, $\mathcal{L}_{\rm SM}$ is the renormalizable SM Lagrangian and $\mathcal{L}_n$ represents higher-dimensional operators of dimension-$n$. In this study, we focus on the dimension-5 and dimension-6 operators only.
In order to ensure the stability of the DM particle, we need to impose a $Z_2$ symmetry, under which the DM field is odd, and the SM fields are even. Therefore, any interaction term involving $\chi$ has to be at least bilinear in this field. Since this bilinear itself has a mass dimension of 3, 
the lowest dimension gauge-invariant operators that can be written down involving the interaction of a Majorana fermion DM and the SM sector are of dimension-5 and involve the Higgs doublet bilinear $H^\dagger H$:

\begin{equation}
\mathcal{L}_5 =  \mathcal{L}_5^{\rm {CPC}}+\mathcal{L}_5^{\rm {CPV}},
\label{dim5}
\end{equation}
where, the CP-conserving operator is 
\begin{equation}
\mathcal{L}_5^{\rm {CPC}} = \frac{g_S}{\Lambda} \overline{\chi} \chi H^\dagger H
\label{dim5cpc}
\end{equation}
and the CP-violating one is given by 
\begin{equation}
\mathcal{L}_5^{\rm {CPV}} = \frac{g_{PS}}{\Lambda} \overline{\chi} i \gamma_5 \chi H^\dagger H.
\label{dim5cpv}
\end{equation}
As mentioned in the introduction, we shall first consider the CP-conserving case, and hence drop $\mathcal{L}_5^{\rm {CPV}}$ from the dimension-5 Lagrangian. Subsequently, we shall discuss the consequences of adding the CP-violating operator. 

After electroweak symmetry breaking, both the operators in Eqs.~\ref{dim5cpc} and $\ref{dim5cpv}$ lead to additional mass terms for the DM field. Moreover, the mass term coming from $\mathcal{L}_5^{\rm {CPV}} $ is purely imaginary. As a result, we obtain a complex mass for the DM field:
\begin{equation}
\mathcal{L}_m= -\frac{1}{2}\overline{\chi} (M_\chi+\frac{g_Sv^2}{\Lambda}+i\frac{g_{PS}v^2}{\Lambda} \gamma_5) \chi.
\end{equation}
However, one can perform the following chiral rotation of the DM field to make the mass term real and positive\footnote{While this manuscript was in preparation, similar observations were made in Ref.~\cite{Kolb-CPC-CPV}}:
\begin{equation}
\chi \rightarrow e^{-i\theta \gamma_5 /2} \chi,
\end{equation}
where,
\begin{equation}
\tan \theta = \frac{g_{PS} v^2}{M_\chi \Lambda + g_S v^2}.
\end{equation}
Such a field redefinition  keeps the kinetic as well as the dimension-6 current-current interactions of the same form, but can of course mix the scalar and pseudo-scalar interactions once CP-violation is allowed. However, since we keep both the co-efficients $g_S$ and $g_{PS}$ arbitrary, such an operator mixing is taken into account. Similarly, in our subsequent analysis, the DM mass is taken as a free parameter, $m_{\chi}$, as the bare mass term $M_{\chi}$ is also a priori arbitrary. 

The complete set of $SU(3)_{\rm C} \times SU(2)_{\rm L} \times U(1)_{\rm Y}$ invariant dimension-6 four-fermion operators are the following:
\begin{align}
\mathcal{L}_6^{4-\rm{Fermi}} & = \frac{1}{\Lambda^2}(\overline{\chi}  \gamma_{\mu} \gamma_5 \chi) \times \sum_{i=1}^3 (g_{LQ} \overline{Q^i_L} \gamma^{\mu}  Q^i_L +g_{Ru} \overline{u^i_R} \gamma^{\mu}  u^i_R +g_{Rd} \overline{d^i_R} \gamma^{\mu}  d^i_R \nonumber \\
                    & +g_{L\ell} \overline{\ell^i_L} \gamma^{\mu}  \ell^i_L +g_{Re} \overline{e^i_R} \gamma^{\mu}  e^i_R ),
\label{4F}
\end{align}
where, the sum over $i$ runs over the three generations, and we have assumed flavour-blind couplings to the SM fermions with the same gauge and global quantum numbers. $Q^i_L$, $u^i_R$, $d^i_R$, $\ell^i_L$ and $e^i_R$ represent the left-handed (LH) quark doublet, right-handed (RH) up-type quark, RH down-type quark, LH lepton doublet and RH lepton fields respectively (the RH fields are singlets under $SU(2)_{\rm L}$). At dimension-6, there is an additional operator involving the derivative of the Higgs field~\cite{Luca,Sacha-Z}
\begin{equation}
\mathcal{L}_6^{\rm Higgs} = \frac{g_D}{\Lambda^2}(\overline{\chi}  \gamma_{\mu} \gamma_5 \chi) (H^\dagger i D^\mu H) + {\rm h.c.}
\label{HiggsDer}
\end{equation}
After EWSB, apart from the $\chi \chi h h Z$ and $\chi \chi h Z$ couplings, $\mathcal{L}_6^{\rm Higgs} $ leads to the following 3-point interaction of $\chi$ with the $Z$-boson:
\begin{equation}
\mathcal{L}_{\overline{\chi}\chi Z} = \frac{g_D g v^2}{2 \cos \theta_W \Lambda^2} (\overline{\chi}  \gamma_{\mu} \gamma_5 \chi) Z^\mu,
\label{ZDM}
\end{equation}
where, $g$ is the $SU(2)_{\rm L}$ gauge coupling, $\theta_{\rm W}$ is the Weinberg angle, and $v=246 \gev$ is the vacuum expectation value of the Higgs field. This term plays a crucial role in determining the low DM mass region in the $m_{\chi}-\Lambda$ plane that satisfies the relic abundance requirement. 

For a Majorana fermion DM, the vector current ($\overline{\chi}  \gamma_{\mu} \chi$) and dipole moments ($\overline{\chi}  \sigma_{\mu \nu} \chi,\overline{\chi}  \sigma_{\mu \nu} \gamma_5 \chi$) vanish identically. While the so-called anapole moment term can exist~\cite{Man}, this operator can be written as a linear combination of the four-fermion operators in Eq.~\ref{4F}, by using the EOM of the gauge field (in this case, the photon field $F_{\mu \nu}$) as follows:
\begin{eqnarray}
\mathcal{L}_{\rm Anapole}  & = & (\overline{\chi}  \gamma_{\mu} \gamma_5 \chi) \partial^\nu F_{\mu \nu} \\
                                           & = & (\overline{\chi}  \gamma_{\mu} \gamma_5 \chi) \sum_f Q_f \left( \overline{f_L} \gamma^\mu f_L +\overline{f_R} \gamma^\mu f_R \right),
\end{eqnarray}
where, $Q_f$ is the electric charge of $f$. Similarly, any operator involving the derivative of the DM field can be eliminated using the EOM of the DM field, which deviates from a free-field equation only with terms proportional to $\Lambda^{-1}$ or lower. 

Since we include only terms upto dimension-6 in the EFT, there is an implicit assumption that all operators are suppressed by a similar scale $\Lambda$, and therefore the contribution of dimension-7 terms in DM phenomenology is sub-leading. For the dimension-7 terms involving the SM fermions, for e.g., $(\overline{\chi} \chi) (\overline{Q_L} H d_R)$ this then readily justifies dropping these operators, since the dimension-6 operators in Eq.~\ref{4F} as well as the Higgs-exchange-induced Yukawa couplings via the scalar operator in Eq.~\ref{dim5cpc} should lead to much stronger interactions with the SM fermions. Although we also drop the dimension-7 couplings with the gauge field strength tensors $G^a_{\mu \nu}$, namely, $\overline{\chi} \chi G^a_{\mu \nu} G^{a \mu \nu}$ (here the gauge field may belong to any of the SM gauge groups), the effective DM coupling to gluons via heavy quark loops is taken into account while considering the spin-independent direct detection rates. Apart from this case, the effect of these couplings is expected to be much smaller than the dimension-5 and 6 terms, since not only are they suppressed by $1/\Lambda^3$, the coupling constant is also of the order of $1/16 \pi^2$, since the DM particle being a gauge singlet, can couple to gauge boson pairs only via loop diagrams.

Generically, an EFT description is valid as long as $\Lambda > > m_{\chi}$, and therefore, it is justified to integrate out the heavy fields with mass of the order of $\Lambda$~\footnote{We adopt the following definitions for the scale $\Lambda$ and the Wilson coefficients for the higher-dimensional operators, $g_i$. $\Lambda$ is assumed to be equal to the mass of the lightest state in the heavy particle sector that is integrated out. Furthermore, if we assume all couplings in the UV completion to be $|g_{\rm UV}|<1$, then it follows that $|g_i|<1$.}. However, in order to get a concrete idea about the minimum value of $\Lambda$ to be considered while computing DM observables, let us take the example of a heavy particle $X$ (scalar or vector) which can mediate in the s-channel annihilation of a DM-pair to a pair of SM particles. An EFT description in this case boils down to replacing  by a constant mass scale the product of couplings of X to the DM-pair ($g_X^{\rm DM}$) and to the SM-pair ($g_X^{\rm SM}$), and its propagator denominator:
\begin{equation}
\frac {g_X^{\rm DM} g_X^{\rm SM}}{s-m_X^2} \rightarrow \frac{-g_X^{\rm DM} g_X^{\rm SM}}{m_X^2} \left(1+\frac{s}{m_X^2}+\mathcal{O}(\frac{s}{m_X^2})^2\right).
\label{EFT-expn}
\end{equation}
In an weakly coupled underlying theory, $g_X^{\rm DM} \sim g_X^{\rm SM} \sim \mathcal{O}(1)$, and henceforth we shall assume this to be the case. Therefore, when matching the UV theory to the EFT, we obtain the relation $\Lambda = m_X$ for $g_X^{\rm DM} \times g_X^{\rm SM}=1$. 
Now, consider the pair annihilation rate of $\chi$ in the early universe, which is relevant for determining its current abundance. In this case, if $v$ is the relative velocity between the DM particles, then the centre of mass energy squared is given by
\begin{equation}
s = 4 m_\chi^2 + m_\chi^2 v^2 + \mathcal{O} (v^4).
\end{equation}
Therefore, the 2nd term in the propagator expansion in Eq.~\ref{EFT-expn} now reads
\begin{equation}
\frac{-s}{m_X^4} \simeq  \frac{-4 m_\chi^2}{m_X^4} - \frac{m_\chi^2 v^2}{m_X^4}.
\end{equation}
Ignoring the $v^2$ piece (which is smaller than the leading term by a factor of $\sim 0.025$), and comparing this term to the leading term of $-1/m_X^2$, we observe that if we assume the minimal requirement of not producing $X$ on-shell in a process, i.e., $m_X > 2 m_\chi$, then the 2nd term in the expansion in Eq.~\ref{EFT-expn} is also of the order of $-1/m_X^2$, which is the same as the leading term. Therefore, in such a case, $m_X > 2 m_\chi$ is a very poor approximation of the full theory. If on the other hand, we assume $m_X > 10 m_\chi$, the 2nd term 
in Eq.~\ref{EFT-expn} can at most be $-1/(25 m_X^2)$, and hence it contributes only $4\%$ of the leading term (at the amplitude level). Therefore, at the level of cross-sections, the error will be less than $8\%$. With such considerations, we find it justified to consider $\Lambda \sim m_X >10 m_\chi$. There can be modifications to this argument if the underlying theory is not weakly coupled or we have a heavy particle mediating in the t-channel etc. Keeping such modifications in mind, while presenting our results, we have separately indicated the regions where $\Lambda > 10 m_\chi$ and where $2 m_\chi < \Lambda < 10 m_\chi$. 

 Apart from the computation of relic density, a careful choice of the scale $\Lambda$ is also necessary for the high-energy collider experiments like LEP and LHC, in order for the EFT to be a valid description. We shall discuss the choice of such scales in Sec.~\ref{sec3}.  For the direct detection experiments, the momentum transfer in the relevant processes is in the MeV scale. Therefore, they do not lead to any further constraint on the region of validity for the EFT. Finally, as far as the indirect detection experiments are concerned, the conditions should be similar to the ones obtained above for the annihilation processes in the early universe. The only difference between DM-pair annihilation rates ($\sigma v$) in the early and the present universe is the DM relative velocity $v$, which was already a small effect in our estimates above for the early universe.

\section{Experimental constraints and the likelihood \\function}
\label{sec3}
In this section, we provide the details of the experimental constraints employed in our analysis: originating from cosmological, astrophysical, laboratory- and collider-based searches. We also briefly describe the profile-likelihood method used in our study and the various uncertainties in the observables that enter into the likelihood. 

\subsection{Profile likelihood method}
\label{PL}
In this study, we employ the profile-likelihood (PL) approach~\cite{PLmethod} to explore high probability regions of the multi-dimensional EFT parameter space. PL is a statistical method motivated in parts by both the Bayesian and the frequentist approaches. It treats the unwanted parameters as nuisance parameters as in the Bayesian theory. However, unlike in a Bayesian approach, in which one marginalizes over all unwanted parameters, the PL method takes the frequentist's concept of  maximizing the likelihood along the directions one is profiling over. In other words, if a model has an $n$-dimensional parameter space, and we are only interested in $p$ of those dimensions, then the PL is obtained by maximizing the likelihood over the $(n-p)$ dimensions we are not interested in. Therefore, unlike in the marginal posterior in Bayesian theory, the prior does not contribute to the PL. However, it is very difficult to cover the full multi-dimensional parameter space with finite samples in a numerical scan. Thanks to the advantage of prior-independence, it is nevertheless possible to combine several fine-grained scans focused on particular regions of the parameter space. For example, even though in a region like $m_W<m_\chi<m_t$,
the DM particle $\chi$ can achieve the required relic density via its pair-annihilation to the $W^+W^-$ final state, this solution spans only a  small volume in the whole hyperspace. A focused scan of such regions, therefore, becomes a necessity.    
 
Our results will be primarily described in the relevant set of two-dimensional parameter regions which are in best agreement with all current experimental data, an example being the ($m_\chi,\Lambda$) space in 68\%~($1\sigma$) and 95\%~($2\sigma$) confidence intervals. After profiling over the rest of the parameters, one can write down confidence intervals in the ($m_\chi,\Lambda$) plane as an integral of the likelihood function $\mathcal{L}(m_\chi,\Lambda)$
\begin{equation}
\frac{\int_{\mathcal{R}}\mathcal{L}(m_\chi,\Lambda)dm_\chi d\Lambda}
{\rm{normalization}}
=\varrho, 
\end{equation}
where, the normalization in the denominator is the total probability with $\mathcal{R}\to\infty$ and $\mathcal{R}$ is the smallest area bound with a fraction $\varrho$ of the total probability. If the likelihood can be modelled as a pure Gaussian distribution, the 68\%~(95\%) confidence intervals in a  two dimensional parameter space corresponds to $-2\ln(\mathcal{L}/\mathcal{L}_{\rm{max}})=2.30~(5.99)$, where $\mathcal{L}_{\rm{max}}$ is the maximum value of the likelihood in the region $\mathcal{R}$. Hereafter, for convenience, we introduce the variable $\chi^2=-2\ln(\mathcal{L})$. We note that 
$\chi^2$ is not exactly the same as in the usual chi-squared analysis unless the likelihood is described by a pure Gaussian. 

The experimental constraints employed in our analysis, along with their central values, $1\sigma$ experimental uncertainties, and functional form of the likelihood functions are shown in Table~\ref{like}. The details of most of the constraints and their likelihoods are provided in the relevant sections referred to in the Table. Our numerical scan of the parameters in the EFT span the following ranges:
\begin{align}
10 \gev \leq \mchi \leq 5 \tev \nonumber ~~\\
2\mchi \leq \Lambda \leq 100 \tev \nonumber \\
-1 \leq g_i  \leq 1. ~~~~~~~~ 
\end{align}
We use a flat prior for all the operator co-efficients in the range $|g_i|<1$. For $\mchi$ and $\Lambda$, we combine both flat and log priors to obtain a maximal coverage of the whole parameter space. As mentioned earlier, the co-efficients of the effective operators are taken to be $|g_i|<1$ , since we assume an weakly coupled UV completion to the EFT.
\begin{table}[t]\footnotesize
\begin{center}
\begin{tabular}{|l|l|l|l|l|l|}
\hline 
Measurement & Central Value & Error ($1\sigma$) & Distribution & Ref.\\
\hline 
Relic density			& $0.1199$ 	& $0.0027$ 		& Gaussian &  \cite{Planck}\\
${\rm BR}(h\to\rm{invisible})$	& $0.0$ 	& $\frac{24\%}{1.64}$		& Gaussian &  \cite{Strumia-inv}\\
$\Gamma(Z\to\rm{invisible}) (\mev)$	& $0.0$ 	& $\frac{2.0}{1.64}$		& Gaussian &  \cite{LEPZwidth}\\
\hline 
\hline
XENON100 $\sigsdn$ (2012)		 & Appendix A 	& Appendix A  		& Gaussian+Poisson &  \cite{XENON}\\
LUX $\sigsip$ (2013)			 &  Appendix A 	& Appendix A 		& Gaussian &  \cite{LUX}\\
Monojet (CMS, 8 TeV, $19.5 {~\rm fb}^{-1}$)	 & Appendix B	& Appendix B  		& Gaussian+Poisson &  \cite{CMS}\\
Mono-photon (LEP, $650 {~\rm pb}^{-1}$)		 & Sec.~\ref{sec:LEP}	& Sec.~\ref{sec:LEP} 		& Gaussian+Poisson &  \cite{LEP-monophoton}\\
Fermi dSphs (5-yrs)	 & Ref.\cite{Sming-1}	& Ref.\cite{Sming-1} 		& Gaussian+Poisson &  \cite{Fermi}\\
IceCube-79	 & Sec.~\ref{ice}		& Sec.~\ref{ice}		& hard cut &  \cite{Ice}\\
\hline 
\end{tabular}\caption{\sl The experimental constraints employed in our analysis, along with their central values, $1\sigma$ experimental uncertainties, and functional form of the likelihood functions. The details of most of the constraints are provided in the relevant sections referred to above.}
\label{like}
\end{center}
\end{table}

\subsection{Relic abundance} 
From the relative heights of the acoustic peaks in the cosmic microwave background, the {\em Planck} experiment has measured the cold dark matter density to an accuracy of $3\%$~\cite{Planck}:
\begin{equation}
\Omega_c h^2 = 0.1196 \pm 0.0031 {~~~~~(68\%  {~\rm C.L., {\it Planck}})}.
\end{equation}
If in addition, the WMAP polarization (WP) data at low multipoles is included, the above number on the $1\sigma$ error changes slightly:
\begin{equation}
\Omega_c h^2 = 0.1199 \pm 0.0027 {~~~~~(68\%  {~\rm C.L., {\it Planck}+WP})}.
\label{eq-relic}
\end{equation}
Although the difference between the two is not very significant, for definiteness, we use the value in Eq.~\ref{eq-relic} in our analysis. The likelihood function is taken as a Gaussian, and apart from the experimental error bar quoted above, an additional theoretical uncertainty of $10\%$ in the computation of $\Omega_{\chi} h^2$ has been assumed. The above number on the relic abundance is taken only as an upper bound, i.e., we demand $\Omega_{\chi} h^2 \leq \Omega_c h^2$,  such that the DM candidate $\chi$ does not overclose the universe. However, we do not assume the existence of some other DM candidate making up for rest of the required relic abundance. Therefore, if for a parameter point $\Omega_{\chi} h^2 < \Omega_c h^2$ as computed within the EFT, then a non-thermal production of $\chi$ should give rise to the additional required DM density. Such a non-thermal mechanism is not described by the EFT, but can exist in the UV completion. For example, in a supersymmetric theory, the late-time decay of gravitinos or moduli fields can produce a neutralino DM. Since such a gravitino or moduli field interacts very feebly with the DM field, it does not affect the DM phenomenology otherwise. {\it To summarize, even though we accept parameter points which satisfy $\Omega_{\chi} h^2 \leq \Omega_c h^2$, the DM particle $\chi$ is assumed to have the relic density of $\Omega_c h^2$ in the present universe, produced with a combination of thermal and non-thermal mechanisms.} 

The relic density~\footnote{While solving the Boltzmann equation for computing the DM relic density, we assume as initial condition an equilibrium thermal abundance of the DM particle.}, as well as all other observables in DM experiments have been computed using the code {\tt micrOMEGAs}~\cite{micro} with the input model files for {\tt CalcHEP}~\cite{calchep} generated using {\tt FeynRules}~\cite{Feynrules}. However, in many cases, we replace the default parameters used in {\tt micrOmegas} for astrophysical and nuclear physics inputs as described in the following subsections.

\subsection{Spin-independent and spin-dependent scattering with nuclei}
\label{DD}
In the non-relativistic limit, relevant for spin-independent (SI) DM scattering with nuclei, the only Majorana DM bilinear that plays a role is the scalar one, $\overline{\chi} \chi$. The pseudo-scalar bilinear $\overline{\chi} \gamma_5 \chi$ in $\mathcal{L}_5^{\rm {CPV}}$ vanishes in the zero DM velocity limit, while the axial vector current $\overline{\chi} \gamma^\mu \gamma_5 \chi$ leads to spin-dependent (SD) scattering. Therefore, only the scalar operator $\mathcal{L}_5^{\rm {CPC}}$ is constrained by the SI scattering limits and the four-fermion interactions with SM quarks as well as the DM interaction with the Z-boson are constrained by the SD limits. For the recent-most bounds on these scattering cross-sections, we have used data from the LUX experiment~\cite{LUX} (for SI) and the XENON100 experiment (for SD)~\cite{XENON}. 

The event rate in direct detection experiments suffers from uncertainties coming from astrophysical and nuclear physics inputs. The astrophysical uncertainties originate from our lack of precise knowledge of the DM local density ($\rho_\odot$) as well as its velocity distribution in the rest frame of the detector, $f(\vec{v}+\vec{v}_E)$, where $\vec{v}$ denotes the DM velocity in the galactic rest frame, and $\vec{v}_E$ represents the motion of the earth with respect to the galactic frame. At present, the experimental collaborations present their limits on SI and SD cross-sections by fixing the astrophysical inputs to specific values, and by assuming $f(\vec{v}+\vec{v}_E)$ to be a truncated Maxwell-Boltzmann distribution, with two additional parameters, the velocity dispersion $\sqrt{\langle v^2 \rangle}$ and the galactic escape velocity $v_{\rm esc}$. In order to take into account the uncertainties of all the astrophysical parameters, we adopt the method developed in Ref.~\cite{Ullio}. Given a choice of the DM density profile in the halo $\rho_\chi(r)$, and a mass model for the Milky Way, Eddington's inversion formula~\cite{Edd} is used in Ref.~\cite{Ullio} to determine the phase-space density function $\rho_\chi(r) f(\vec{v},t)$. The two primary assumptions in this approach are that the DM particle $\chi$ makes up the entire DM component of the Universe, and that the DM distribution is spherically symmetric. We adopt the phase-space density factor and its associated $2\sigma$ error bars as computed in Ref.~\cite{Ullio}, to which we refer the reader for further details~\footnote{The required data files are provided by the authors of Ref.~\cite{Ullio} in an electronic format in \tt{http://arxiv.org/src/1111.3556v2/anc}}. The Burkert profile is chosen for $\rho_\chi(r)$, which tends to give a slightly larger velocity dispersion compared to the NFW and Einasto profiles~\cite{Ullio}.

The nuclear physics uncertainties in SI and SD scattering rates stem from the corresponding nuclear matrix elements $\langle N| \bar{q}q |N\rangle$ and $\langle N| \bar{q}\gamma_\mu\gamma_5 q |N\rangle$ respectively. Recent progress in lattice QCD calculations predict a rather small value for the strange quark content of the nucleon, $f_{Ts}$, and the results from different lattice simulation groups seem to have converged on this fact. Similarly, the pion-nucleon sigma term $\Sigma_{\pi N}$, entering the SI rates along with $f_{Ts}$,  has also been determined by lattice calculations rather precisely. For SD scattering rates, the nuclear physics inputs are encoded by $\Delta q^n$ ($q=u,d,s$), which gives the fraction of spin due to each quark in the neutron (the corresponding numbers for proton are related by isospin rotation). The values of these nuclear physics inputs used in our calculations, along with their uncertainties, are listed in Table~\ref{tab:nucli_params}.

\begin{table}
\begin{center}
\begin{tabular}{|l| l| l|}
\hline
\hline
\multicolumn{3}{|c|}{Hadronic nuisance parameters} \\
\hline\hline
$\Sigma_{\pi N}$ & $ 41\pm 6$ \mev & \cite {Alvarez-Ruso:2013fza}\\
$f_{Ts}$ & $0.043\pm 0.011$  & \cite{Junnarkar:2013ac}\\
$\Delta u^n$ & $ -0.319 \pm 0.066 $ &  \cite{QCDSF:2011aa} \\
$\Delta d^n$ & $ 0.787 \pm 0.158 $ & \cite{QCDSF:2011aa}\\
$\Delta s^n$ & $ -0.020 \pm 0.011 $ & \cite{QCDSF:2011aa}\\
\hline
\hline
\end{tabular}
\end{center}
\caption{\sl Nuclear physics inputs, used in this study for computing SI and SD DM scattering rates with nuclei, as determined by latest lattice simulations.}
\label{tab:nucli_params}
\end{table}
For further details on the construction of the likelihood function for LUX (SI) and XENON100 (SD), we refer the reader to Appendix-A. 

\subsection{DM annihilation in the present universe}
Pair annihilation of DM particles in the present universe can lead to several SM final states. In our analysis, we include constraints from gamma ray and neutrino searches. Since the constraints obtained from the observation of gamma rays originating at Milky Way satellite galaxies are more robust than those obtained from Galactic Centre observations, we consider only the former~\footnote{For a recent discussion of constraints coming from the Galactic Centre observations, see, for example, Ref.~\cite{Kelso}. However, these constraints become weaker compared to the dSphs ones, if one adopts a more cored profile for the DM halo, an example being the Burkert profile~\cite{Salucci}.}.

In this connection, it should be mentioned that DM pair annihilations can also lead to positron and anti-proton signals. However, astrophysical positron backgrounds are not yet known precisely enough to use them as robust constraints. As for anti-protons, although constraints can be obtained, but they depend strongly on the used propagation model for anti-protons under the galactic magnetic fields. Since we want to determine as robust a limit on the parameter space as possible, even though relevant in certain cases, we abstain from using the positron and anti-proton data in this study.

\subsubsection{Gamma ray observations}
In the EFT setup considered here, the gamma ray line signal can either appear from loop-level processes involving SM fermions, or from dimension-7 operators which are suppressed for reasons explained in Sec.~\ref{sec2}. Therefore, as far as indirect detection signals are concerned, the continuum gamma ray  observations can put constraints on our scenario, although the scalar and axial-vector DM currents lead to annihilation rates in the present Universe which are p-wave suppressed. Hence the gamma ray observations are mostly relevant for the pseudo-scalar operator in our EFT setup. 

We use the constraints obtained by Fermi Large Area Telescope (Fermi-LAT) observation of diffuse gamma rays from the milky way satellite galaxies (dwarf spheroidal galaxies (dSphs))~\cite{Fermi} by including the eight classical dSphs in our analysis, since the dark matter distribution in the classical dSphs is measured with a higher accuracy from the velocity dispersion of the luminous matter~\cite{Trotta}. We combine the 273 weeks' Fermi-LAT data ( from 2008-08-04 to 2013-10-27) using the  Pass-7 photon selection criterion, as implemented in the Fermi Tools. The J-factors for the dSphs included are taken from Table-I in Ref.~\cite{Fermi}, where we have used the numbers corresponding to the NFW profile. If instead the Burkert profile is used, the central values of the J-factors are very similar, while the uncertainty bands in NFW are slightly wider. Our method for combining the likelihood function corresponding to different energy bins is the same as that used in the original Fermi-LAT analysis. However, we predetermine, in a model independent manner, the likelihood map in the residual flux - energy plane, by combining the data for the eight classical dSphs. Here, the residual flux refers to the background subtracted gamma ray flux, scaled by the J-factor. For details on the statistical analysis, we refer the reader to Ref.~\cite{Sming-1}~\footnote{We thank Qiang Yuan for providing us the likelihood map using the 273 weeks' data, and Xiaoyuan Huang for careful cross-checks.}. We include data in the whole energy range as observed by Fermi, namely, 200 MeV to 500 GeV. In the dSphs likelihood function used by us, the J-factor uncertainty is modelled by a Gaussian function, while the statistical uncertainty in the number of observed photons is modelled by a Poission distribution, with parameters as given in Ref.~\cite{Sming-1}. 

\subsubsection{Neutrino telescopes}
\label{ice}
The IceCube neutrino telescope has been looking for muon neutrinos originating from DM annihilations in the Sun. The 317 days' data collected with the 79-string IceCube detector (including the DeepCore subarray) during the period June 2010 to May 2011 is found to be consistent with the expected atmospheric backgrounds~\cite{Ice}, and therefore leads to bounds on DM annihilation rates in the Sun to final states involving neutrinos. Assuming that the DM capture and annihilation rates in the Sun are in equilibrium, these bounds can then be translated to limits on SI and SD scattering cross-sections of the DM particles with proton ($\sigma_{p-\chi}^{\rm SD}$). While for SI scattering rates the XENON100 and LUX constraints are much stronger, for $\sigma_{p-\chi}^{\rm SD}$ scattering with $\mchi>35$ GeV, the IceCube limits are stringent and competitive with ground-based experiments. We should remark here that, due to the higher spin expectation value of the neutron group compared to the proton group in XENON nuclei, the bounds on $\sigma_{n-\chi}^{\rm SD}$ from XENON100 is stronger than the bound on $\sigma_{p-\chi}^{\rm SD}$. This is the reason the IceCube limits on $\sigma_{p-\chi}^{\rm SD}$ are important even after including the XENON100 limits on SD scattering rates, since in the EFT, the $\sigma_{p-\chi}^{\rm SD}$ can be enhanced compared to $\sigma_{n-\chi}^{\rm SD}$ in certain regions of the parameter space (for example, with $|g_{LQ}-g_{Ru}|> |g_{LQ}-g_{Rd}|$).

As is well-known, the energy spectrum of neutrinos produced from the annihilation channel $\chi \chi \rightarrow W^+ W^-$ (or, $\chi \chi \rightarrow \tau^+ \tau^-$ for $\mchi < M_W$) is harder than that produced from the channel $\chi \chi \rightarrow b \bar{b}$. Hence, the former leads to stronger bounds on $\sigma_{p-\chi}^{\rm SD}$, assuming a $100\%$ annihilation to either of the channels. Since the limit from the $b \bar{b}$ channel is the weakest, any parameter point in the EFT, with any branching ratio to the $b\bar{b}$ final state, has to at least satisfy this limit.  Therefore, in our analysis, we reject parameter points which lead to $\sigma_{p-\chi}^{\rm SD}$ higher than the $95 \%$ C.L. IceCube bound as obtained assuming the $b \bar{b}$ mode of DM annihilation.

\subsection{Collider search: LEP}

\subsubsection{Z-boson invisible width}
For $m_{\chi} < M_Z/2$, the interaction term in Eq.~\ref{ZDM} will lead to the invisible decay $Z \rightarrow \chi \chi$. The decay width of the Z-boson has been precisely measured at the LEP experiment, and apart from the width originating from $Z \rightarrow \nu \bar{\nu}$, the $95\%$ C.L. upper bound on the invisible width of the Z-boson is given by~\cite{LEPZwidth}
\begin{equation}
\Gamma^Z_{\rm inv} < 2 {~\rm MeV}~~~(95\% {~\rm C.L.}, ~{\rm LEP}).
\end{equation}
This, therefore, acts as a powerful constraint on the dimension-6 operator in Eq.~\ref{HiggsDer} for a light DM. The likelihood function is taken as a Gaussian with parameters as shown in Table~\ref{like}.

\subsubsection{Mono-photon search}
\label{sec:LEP}
Single photon events were looked for at the LEP collider to search for signatures of graviton production, the null results of which leads to bounds on the radiative process $e^+ e^- \rightarrow \chi \chi \gamma$. In this paper, we use the limits from the DELPHI collaboration obtained using the $650 {~\rm pb}^{-1}$ of LEP2 data with centre of mass energy in the range $180-209$ GeV~\cite{LEP-monophoton}. We compute the relevant cross-sections using {\tt MadGraph5}~\cite{MG5}, with the model files generated using {\tt FeynRules}~\cite{Feynrules}, and use our own detector simulation code to model the DELPHI detector response. The DELPHI results on the dominant SM background process $e^+ e^- \rightarrow \nu_\ell \overline{\nu_\ell} \gamma$ are reproduced by our Monte Carlo (MC) simulation to a reasonably good accuracy, which is then used to compute the signal predictions in the EFT framework. The LEP data in the monophoton channel affects the DM couplings to the Z-boson and to the charged leptons, (i.e., the couplings $g_D, g_{L_L}$ and $g_{R_E}$). The likelihood function is a convolution of Poission and Gaussian distributions, and is of the same form as the LHC monojet search likelihood described in Appendix B (Eq.~\ref{LHC-like}). The expected number of background events as well as the observed number of events after the cuts are taken from the DELPHI results~\cite{LEP-monophoton}.

Since we impose the condition $\Lambda > 10 \mchi$ from relic density considerations, we have, for $\sqrt{s}=200$ GeV, $\Lambda > \sqrt{s}$, as long as $\mchi>20$ GeV. For DM mass in the range $10-20$ GeV, the EFT results can overestimate the production cross-section, and bounds on $\Lambda<200$ GeV may not be valid. However, we estimated the expected cross-sections using $s-$ and $t-$ channel exchange of a mediator of mass $M$ as well, and compared those to the predictions of the EFT with a scale $\Lambda \sim M$. Even on taking this cross-section overestimate into account, for low mass DM in the window of $10 {~\rm GeV} < m_\chi < 20 {~\rm GeV}$, due to very large production rates, the corresponding UV completions with $s-$  and $t-$ channel mediator exchange should be excluded as well.

\subsection{Collider search: LHC}

\subsubsection{Higgs boson invisible width}
The properties of the Higgs-like boson have now been measured by the ATLAS and CMS experiments in different search channels with varying degrees of accuracy~\cite{Higgs-disco}. Although the direct search for a Higgs boson decaying invisibly is not yet very sensitive, global fits to the Higgs data, assuming that the production cross-sections and partial decay width of the Higgs in all other channels are the same as in the SM, lead to the following upper bound on the Higgs boson invisible branching ratio~\cite{Strumia-inv}
\begin{equation}
{\rm BR}(h \rightarrow \chi \chi) = \frac{\Gamma(h \rightarrow \chi \chi)} {\Gamma^{h^{\rm SM}}_{\rm Total}+\Gamma(h \rightarrow \chi \chi)} 
                                                 < 0.24 ~~~(95\% {~\rm C.L.}, ~{\rm LHC})
\end{equation}
We have used $\Gamma^{h^{\rm SM}}_{\rm Total} = 4.21$ MeV for $m_h=126.0$ GeV in our numerical analysis. The Higgs invisible search constraints the CP-even and -odd scalar operators considerably, in the DM mass range in which the decay mode is kinematically allowed.

\subsubsection{Monojet plus missing energy search }
\label{LHC}
The ATLAS and CMS collaborations have searched for events with at least one energetic jet and missing transverse momentum ($\met$) in the 7 and 8 TeV LHC data. In $pp$ collisions, DM can be pair produced if it has effective interactions with quarks and gluons. Since the DM particles themselves are invisible, the events are triggered by the presence of at least one hadronic jet, against which the DM pair recoils, giving rise to the $\met$ in the events. In this paper, we adopt the CMS analysis with $19.5 \fb$ of data collected at the 8 TeV centre of mass energy~\cite{CMS}. 

The operator basis constrained by CMS is not written in an $SU(2)_L \times U(1)_Y$ invariant form, and therefore differs from ours. Moreover, as discussed before, we consider the presence of all operators at the same time. This requires us to perform the Monte Carlo (MC) analysis following CMS within our EFT framework. In order to validate our MC, we first reproduce the CMS results, and compare the $95\%$ C.L. upper bounds in the $m_\chi - \Lambda$ plane for the axial vector current-current four fermi interaction with quarks. The agreement found is accurate to within $5\%$. We briefly discuss this validation procedure and our MC setup in Appendix-B, where the LHC likelihood function is also described.

The validity of an EFT approach at a high-energy hadron collider like the LHC has recently been discussed at length~\cite{validity-dolan, validity-riotto}. Since rather strong jet $p_T$ and $\met$ cuts are used by the collaborations in order to reduce the very large SM backgrounds, the subprocess centre of mass energy involved is also large. In such a case, if the suppression scale of the operators $\Lambda$ (or equivalently the mediator mass in a weakly coupled UV completion) is comparable to the subprocess COM energy, the mediator particles can be produced on-shell, and an EFT description breaks down. As an order of magnitude estimate, one can consider the minimum value of the cut-off scale, in order for the EFT to be valid as~\cite{validity-dolan}

\begin{equation}
\Lambda > \sqrt{4 m_\chi^2 + (\met_{\rm min})^2},
\end{equation}
where, for the CMS analysis, $\met_{\rm min}=400$ GeV. For low DM masses, the above requirement is dominated by the $\met$ cut. Comparing this order of magnitude bound to the condition imposed by us ($\Lambda > 10 m_\chi$), we can see that, roughly for $m_\chi > 41$ GeV, we can safely use the EFT framework at the LHC when $\Lambda > 10 m_\chi$. For $10 \gev < \mchi <40 \gev$, we find that even if $\Lambda>10 \mchi$, the expected cross-section after the CMS cuts is so large that even if the EFT limit is stronger than that given by an UV complete theory including the mediator particles, one can still exclude these points in the UV theory as well.

\section{Relic density: the role of individual operators}
\label{relic-role}
\begin{figure}[h!]
\begin{center}
\centerline{\epsfig{file=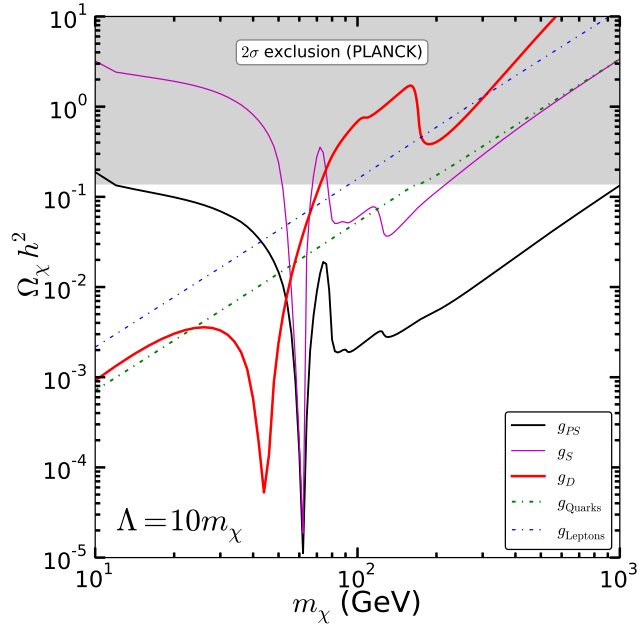,width=9cm,height=9cm,angle=-0,keepaspectratio}}
\caption{\small \sl The thermal component of the relic abundance $\Omega_{\chi} h^2$ as a function of the DM mass $\mchi$, for $\Lambda=10 \mchi$ and one particular coefficient $g_i=1$, the rest of them being set to zero. The contribution of each operator is shown separately for illustration only. The grey shaded region is excluded by the {\it Planck} constraint in Eq.~\ref{eq-relic} at $95\%$ C.L. (for one parameter, $\mchi$).}
\label{relic}
\end{center}
\end{figure}

The requirement that $\Omega_{\chi} h^2 {~\rm(Thermal)} \leq \Omega_c h^2$, places important lower bounds on the operator coefficients $g_i/\Lambda$ and $g_i/\Lambda^2$. Although all the couplings enter together into the computation of the thermally averaged annihilation cross-section in the early universe ($\sv$), there are specific regions of the DM mass where some of them play a dominant role. We illustrate the role of each operator separately in determining $\Omega_{\chi} h^2$ in Fig.~\ref{relic} as a function of $\mchi$ for $\Lambda=10 \mchi$ and the corresponding operator coefficient $g_i=1$. The grey shaded region is excluded by the {\it Planck} constraint in Eq.~\ref{eq-relic} at $95\%$ C.L. (for one parameter, $\mchi$). 
For certain values of $\mchi$ a very low $\Omega_{\chi} h^2$ is observed in Fig.~\ref{relic}. This essentially implies that for those $\mchi$ the requirement on $\Omega_{\chi} h^2$ can be satisfied even for very large values of $\Lambda$ or for very small values of the coupling constants $g_i$. We now briefly discuss the case for each operator separately.

\subsection {Dimension-5 Higgs portal operators} 
With only the CP-conserving Higgs-portal operator $\mathcal{L}_5^{\rm {CPC}}$ it is possible to satisfy the relic density requirement in some specific regions.
For $\mchi \sim M_h/2$, the s-channel Higgs exchange diagram for DM annihilation to light fermion final states (including upto bottom quarks) is resonantly enhanced and we find a sharply peaked region in $\mchi$ where $\Omega_{\chi} h^2$ is below around $0.15$. Since the $126 \gev$ SM-like Higgs boson has a very small SM width ($4.21$ MeV), even on inclusion of an additional invisible width to the DM pair, the Higgs resonance region is very sharply peaked.

Below the Higgs resonance region, for $\mchi \lesssim 50 \gev$, the scalar operator alone leads to very low $\sv$, since the s-wave term in $\sv$ is suppressed by $m_f^2/M_h^2$ (the dominant mode being annihilation to $b\bar{b}$), and the p-wave contribution is not sufficiently large. For the pseudo-scalar operator $\mathcal{L}_5^{\rm {CPV}}$, the helicity suppression in the s-channel diagram is lifted, and the s-wave piece in $\sv$ is now proportional to $\mchi^2/M_h^2$. Therefore, with $\mathcal{L}_5^{\rm {CPV}}$, it is possible to satisfy the relic density criterion even for $\mchi < 50 \gev$ via annihilation to light fermion final states, as seen in Fig.~\ref{relic}. 

For $\mchi \gtrsim 80 \gev$ the annihilation mode $\chi \chi \rightarrow WW$ and for $\mchi \gtrsim 173 \gev$, the mode $\chi \chi \rightarrow t \bar{t}$ open up, leading to the minimum required $\sv$. The first of these annihilation modes receives contribution only from the dimension-5 Higgs portal operators, while the latter one also receives contributions from the dimension-6 four-fermion operator. The s-channel Higgs exchange diagram to the $t \bar{t}$ final state is not p-wave suppressed for the scalar operator, since the s-wave contribution proportional to $m_t^2$ is large enough. With the operator $\mathcal{L}_5^{\rm {CPC}}$, it is possible to obtain the required $\sv$ for $\mchi \lesssim 200 \gev$ (with $\Lambda=10\mchi$). In the CP-violating case, due to the additional s-wave contributions coming from all other light fermion annihilation channels, we can have the required $\sv$ for DM masses upto $1 \tev$.

\subsection{Dimension-6 four-fermion operators}

In Fig~\ref{relic} we also see the role of the dimension-6 four-fermion operators (Eq.~\ref{4F}) in determining $\Omega_{\chi} h^2 $. The blue dot-dashed line shows the contribution of the leptonic operators with $g_{L\ell}=g_{Re}=1$ for all three fermion generations, while the green dot-dashed line shows the same for the operators involving the quarks with $g_{LQ}=g_{Ru} =g_{Rd}=1$. For a given $\mchi$ and $\Lambda$, the quark operators in general lead to a higher $\sv$, and therefore lower $\Omega_{\chi} h^2 $, mainly because of the additional colour factor in the amplitudes. With the leptonic operators the required $\sv$ can be achieved for $\mchi \lesssim 100 \gev$, and with the quark operators for $\mchi \lesssim 180 \gev$ (with $\Lambda=10\mchi$)~\footnote{An additional large s-wave contribution to $\sv $, and therefore a drop in $\Omega_{\chi} h^2 $,  is expected after the $t \bar{t}$ annihilation mode opens up. However, this is not visible in Fig.~\ref{relic}, since for this figure we have set  $g_{LQ}=g_{Ru}=1$, and therefore only the vector current $\bar{t} \gamma^\mu t$ survives, which does not lead to an $m_t^2$ enhanced term in $\sv$. When $g_{LQ} \neq g_{Ru}$ such a feature is observable due to the additional axial vector current, see, for e.g., Fig.~\ref{CPC_mx_lam} (left panel).}.

\subsection{Dimension-6 Higgs derivative operator}
As we have seen in Sec.~\ref{sec2}, the dimension-6 operator $\mathcal{L}_6^{\rm Higgs}$ (Eq.~\ref{HiggsDer}) involving the derivative of the Higgs field leads to the crucial interaction of the DM field with the $Z$ boson in Eq.~\ref{ZDM}. This interaction alone is sufficient to obtain the required annihilation rate $\sv$ in the mass range $10 \gev \lesssim \mchi \lesssim 70 \gev$ with $\Lambda=10 \mchi$ as seen in Fig.~\ref{relic}. Around $M_{\chi}= 45 \gev ~(=M_Z/2)$, there is a resonant enhancement of $\sv$. This resonance is sufficiently broad compared to the Higgs-resonance described above, owing to the much larger Z-boson width ($2.5 \gev$).  

\section{Results: CP-conserving scenario}
\label{sec4}
We discuss the results of the likelihood analysis first in the CP-conserving case, whereby the operator $\mathcal{L}_5^{\rm {CPV}}$ in Eq.~\ref{dim5cpv} is dropped. Subsequently we consider the CP-violating case including $\mathcal{L}_5^{\rm {CPV}}$. Such a separation of the two cases can be justified by the fact that the underlying UV completion of the EFT can be a theory where the scalar sector conserves CP. 

\subsection{Allowed region in $\mchi-\Lambda$ plane}
\label{CPC_all}
\begin{figure}[h!]
\begin{center}
\centerline{\epsfig{file=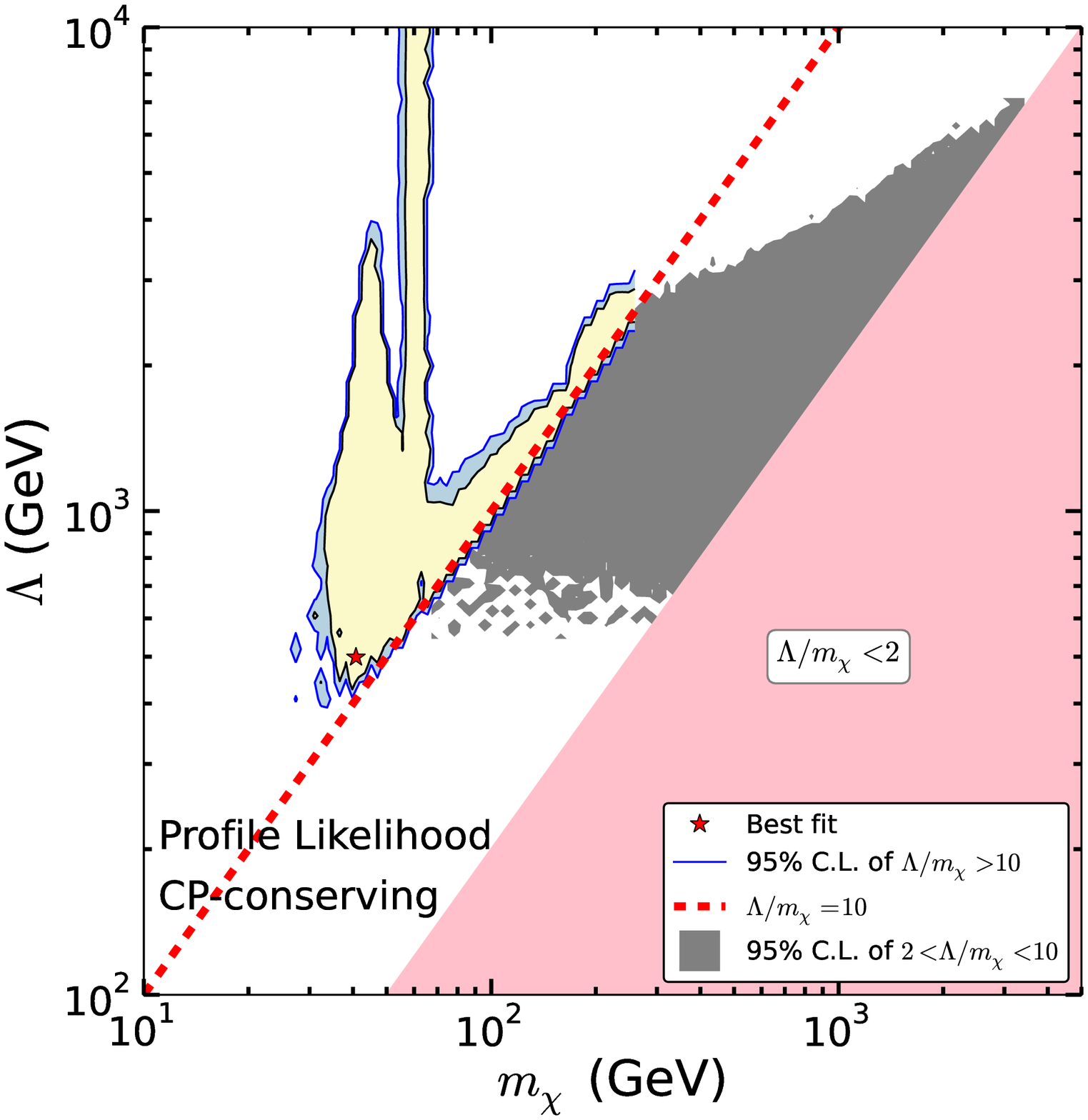
,width=8.4cm,height=8.4cm,angle=-0,keepaspectratio}\
\epsfig{file=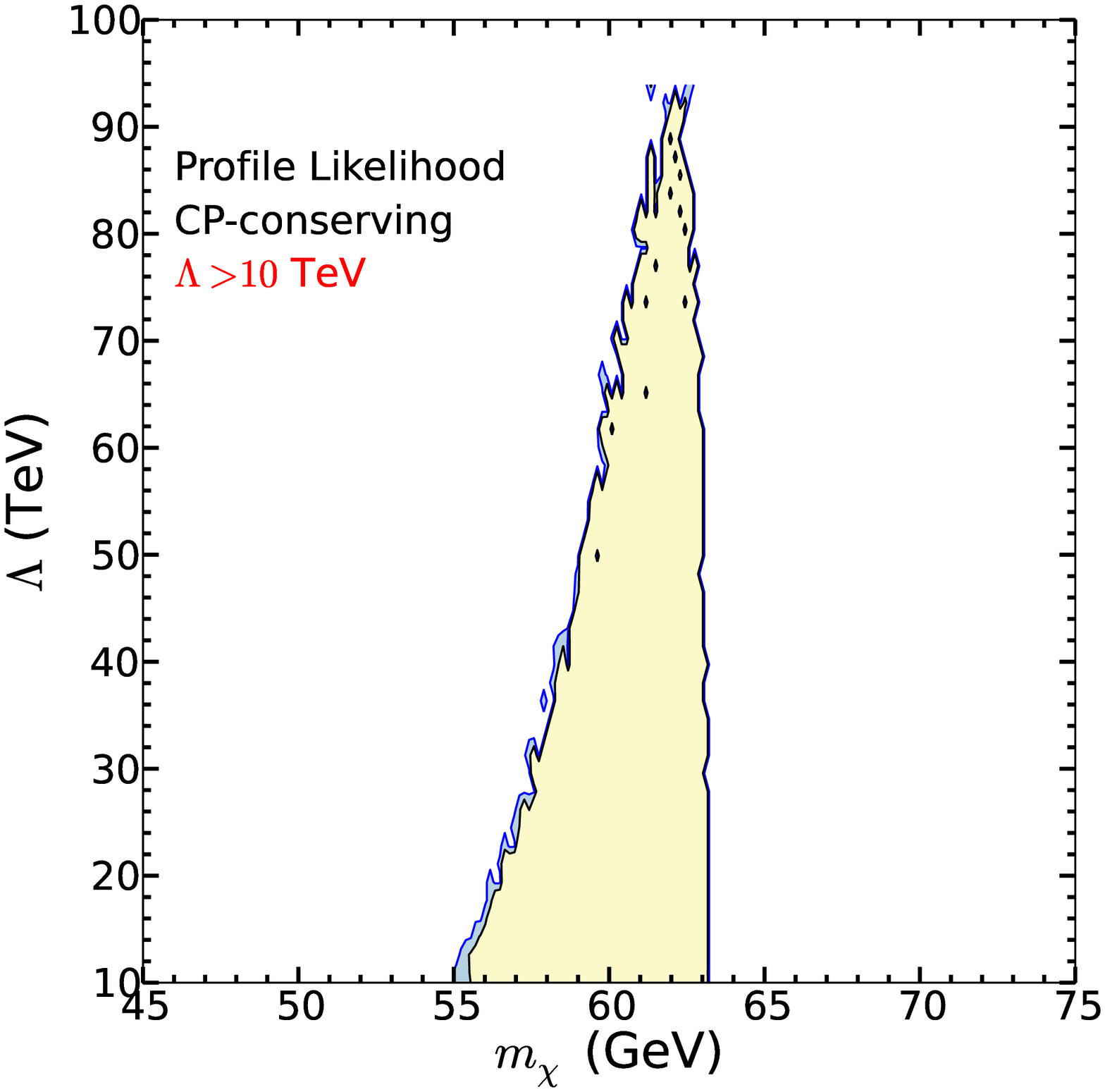
,width=8.4cm,height=8.4cm,angle=-0,keepaspectratio}}
\caption{\small \sl Left panel: $68\%$ (yellow) and $95\%$ C.L. (blue) allowed regions in the $m_\chi-\Lambda$ plane with $\Lambda > 10 m_\chi$. For completeness, the $95\%$ C.L. (grey) allowed region with $2 \mchi <\Lambda < 10 m_\chi$ is also shown, although the validity of the EFT in computing relic abundance and collider observables in this region is questionable. The EFT description is not valid for $\Lambda < 2 \mchi$, which is shown in pink. Right panel: Allowed parameter space near the Higgs resonance region for $\Lambda > 10 \tev$.}
\label{CPC_mx_lam}
\end{center}
\end{figure}

In the EFT analysis including all operators upto dimension-6 simultaneously, the most important two-dimensional likelihood map is in the $m_\chi-\Lambda$ plane. In Fig.~\ref{CPC_mx_lam} (left panel) we show the $68\%$ (yellow) and $95\%$ C.L. (blue) allowed regions in the $m_\chi-\Lambda$ plane, whereby the condition $\Lambda > 10 m_\chi$ has been imposed. For completeness, we also show the $95\%$ C.L. (grey) allowed region with $2 \mchi <\Lambda < 10 m_\chi$, although as discussed before, in this region the validity of the EFT from the point of view of relic abundance and collider physics computations is questionable.  Finally, the region with $\Lambda < 2 \mchi$, where an EFT analysis cannot be applied, is shown in pink. As mentioned in Sec.~\ref{PL}, the two-dimensional allowed regions are obtained after profiling out rest of the parameters in the EFT, which, in this case, are the $\mathcal{O}(1)$ coefficients $g_{i}$ for the different operators.

The shape of the allowed contours is mostly determined by the requirement that $\Omega_{\chi} h^2 ~({\rm Thermal}) < \Omega_c h^2 ~({\rm Planck})$ as discussed in detail in the previous subsection. We now discuss the allowed and ruled out regions of DM mass $\mchi$ by dividing them in a few windows:
\begin{itemize}
\item $\mchi \lesssim 30 \gev$: This region of DM mass is mostly ruled out at $95\%$ C.L. The dimension-6 four-fermion couplings with leptons are excluded for these low DM masses by the LEP mono-photon search, while the couplings with quarks are ruled out by the LHC monojet search. The DM coupling with the Z-boson for this mass range leads to a considerable Z-boson invisible width, and is therefore constrained by the LEP measurements. The latter coupling is also constrained by the LEP mono-photon search. As seen in the previous subsection, the dimension-5 scalar coupling with the Higgs boson cannot furnish a correct relic abundance in this mass range. 

\item $30 \gev \lesssim \mchi \lesssim 50 \gev$: The allowed points in this region are dominated by the Z-resonance in DM annihilation. Since the $\overline{\chi} \chi Z$ coupling is rather weakly constrained by SD direct detection experiments, and near the Z-resonance the $Z$ invisible width constraint also becomes weaker, there is a significant area in the $\mchi-\Lambda$ plane which escapes all current constraints while satisfying the $\Omega_{\chi} h^2$ requirement. The LEP mono-photon search also looses its constraining power on $\mathcal{L}_{\overline{\chi} \chi Z}$ beyond $\mchi \simeq 30 \gev$ for $\Lambda > 10 \mchi$. Since the coupling with the Z-boson does not lead to SI scattering with nuclei, and the bound on SD scattering rates are much weaker, this is one of the DM mass ranges hard to probe with current experiments.

\item $50 \gev \lesssim \mchi \lesssim 70 \gev$: Here, the Higgs-resonance in DM pair-annihilation dominates $\sv$. The Higgs resonance region is not completely seen in Fig.~\ref{CPC_mx_lam} (left panel), which shows values of $\Lambda$ only upto $10 \tev$. We show the $\Lambda>10 \tev$ region separately in the right panel of Fig.~\ref{CPC_mx_lam}, in which the resonance region spans $55 \gev \lesssim \mchi \lesssim 63 \gev$. It should be noted that the thermal averaging integral involved in computing $\sv$ fluctuates in the numerical routine used (as implemented in {\tt micrOMEGAs}), due to which the highest value of $\Lambda$ allowed ($\sim 94 \tev$) is subject to some uncertainty. For such high values of $\Lambda$, $g_S/\Lambda = \mathcal {O}(10^{-5} ~{\rm GeV}^{-1})$, and the spin-independent scattering rate with nuclei $\sigma_p^{\rm SI} \sim 10^{-12}$ pb, while the latest LUX limits can only constrain $\sigma_p^{\rm SI} \sim 10^{-9}$ pb in this range of $\mchi$. Therefore, this is another DM mass region which is not completely probed by experiments so far. 

\item $70 \gev \lesssim \mchi \lesssim 250 \gev$: This allowed region is away from any resonance, and all the operators contribute to $\sv$. Only values of $\Lambda$ very close to the allowed minimum value of $10 \mchi$ are seen to be viable. For $\mchi \gtrsim m_t$, as discussed in Sec.~\ref{relic-role}, there are additional contributions to $\sv$ proportional to $m_t^2$, and therefore slightly larger values of $\Lambda$ are also allowed. 
\end{itemize}

It is interesting to note that, when the condition $\Lambda>10 \mchi$ is imposed, there is an upper limit on the allowed value of $\mchi$, mainly coming from the relic density constraint. For $\Lambda>10 \mchi$, this upper limit is $\mchi \simeq 300 \gev$, while for $\Lambda>2 \mchi$, it extends upto $\mchi \simeq 3 \tev$, as can be seen in Fig.~\ref{CPC_mx_lam}.

\subsection{Future prospects}
\label{sec:CPC-fut}
Having determined the currently allowed parameter space for the CP-conserving scenario in the $\mchi-\Lambda$ plane in the previous subsection, we now discuss the predictions for future experiments in this region. We focus on the future direct detection experiments (both SI and SD), as well as an $e^+ e^-$ collider experiment like the proposed International Linear Collider (ILC)~\cite{ILC}. In the CP-conserving case, the predictions for DM annihilation rates in the present Universe, $\langle \sigma v \rangle_0$, are rather low in our EFT framework (below $10^{-27} ~{\rm cm}^3~{\rm sec}^{-1}$), since most of the operators lead to annihilations which are p-wave suppressed. Therefore, the reach of future gamma-ray observation experiments is not very high in this case, except for $\mchi > m_t$, where one can obtain a rate as high as $\langle \sigma v \rangle_0 \sim \mathcal{O}(10^{-26}) {~\rm cm}^3 {~\rm s}^{-1}$ due to the additional s-wave contribution.

\begin{figure}[htb!]
\centering
\centerline{\includegraphics[width=8.5cm,height=8.5cm,keepaspectratio]{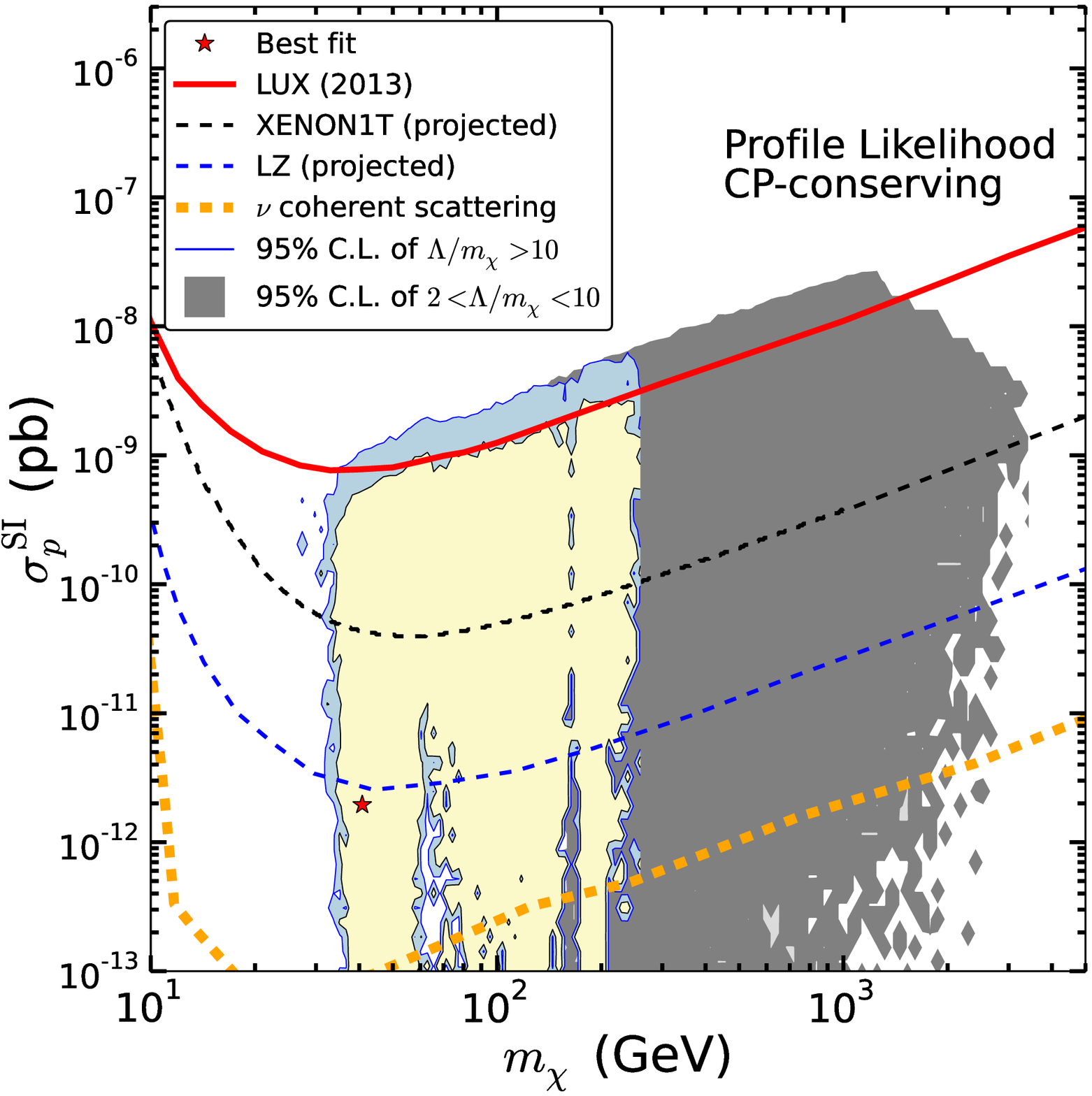}\
\includegraphics[width=8.5cm,height=8.5cm,keepaspectratio]{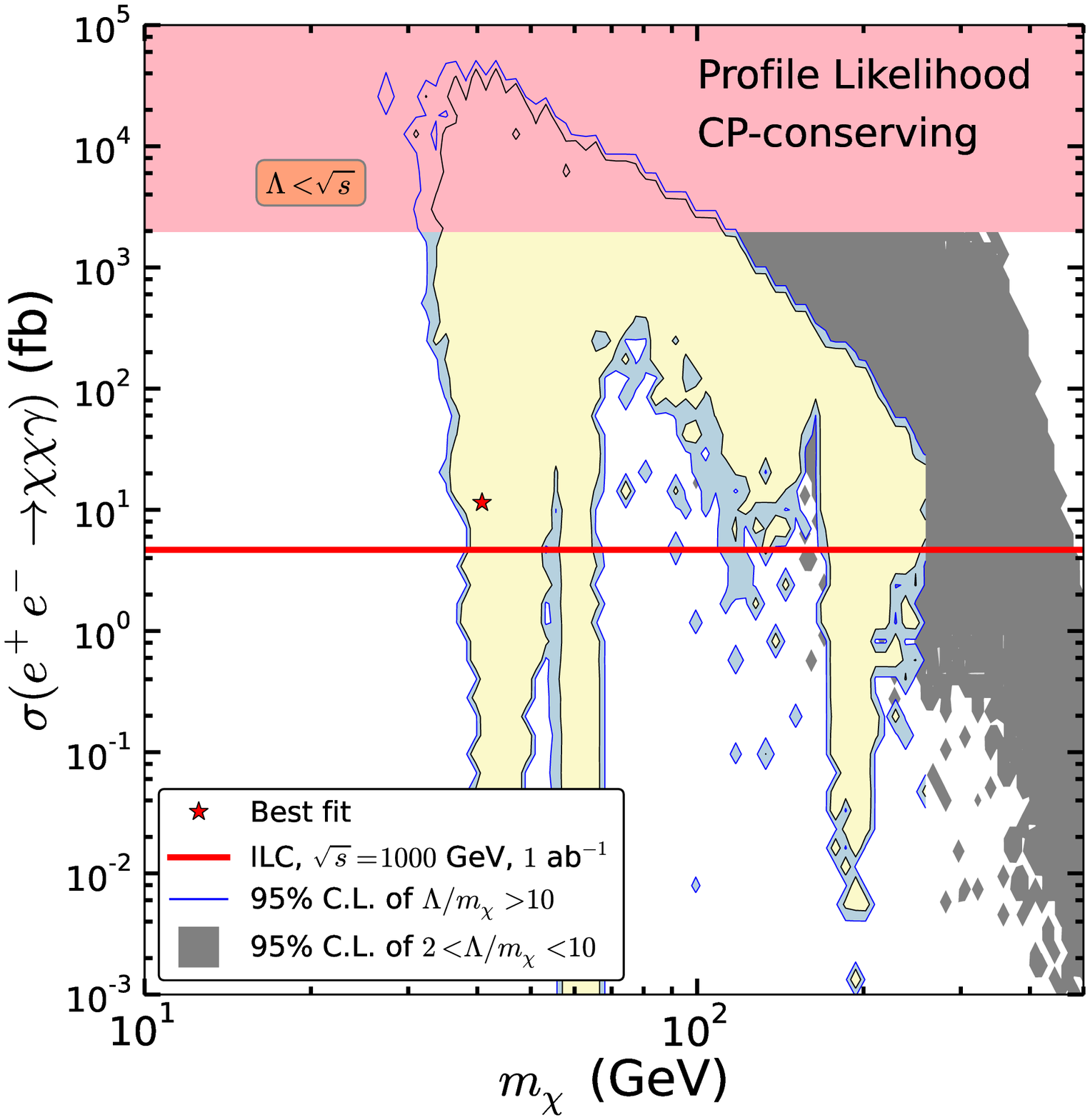}}
\centerline{\includegraphics[width=8.5cm,height=8.5cm,keepaspectratio]{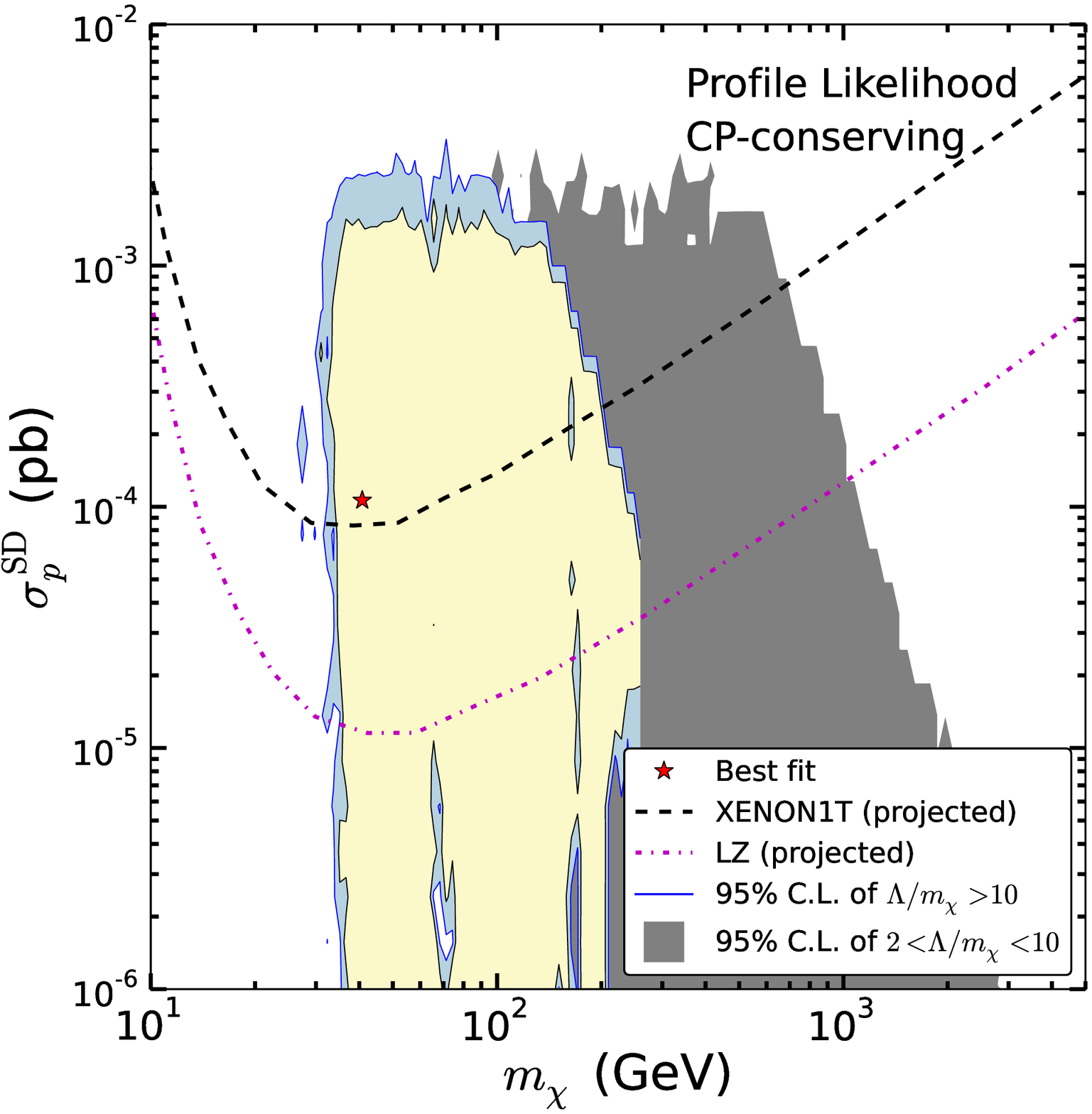}\
\includegraphics[width=8.5cm,height=8.5cm,keepaspectratio]{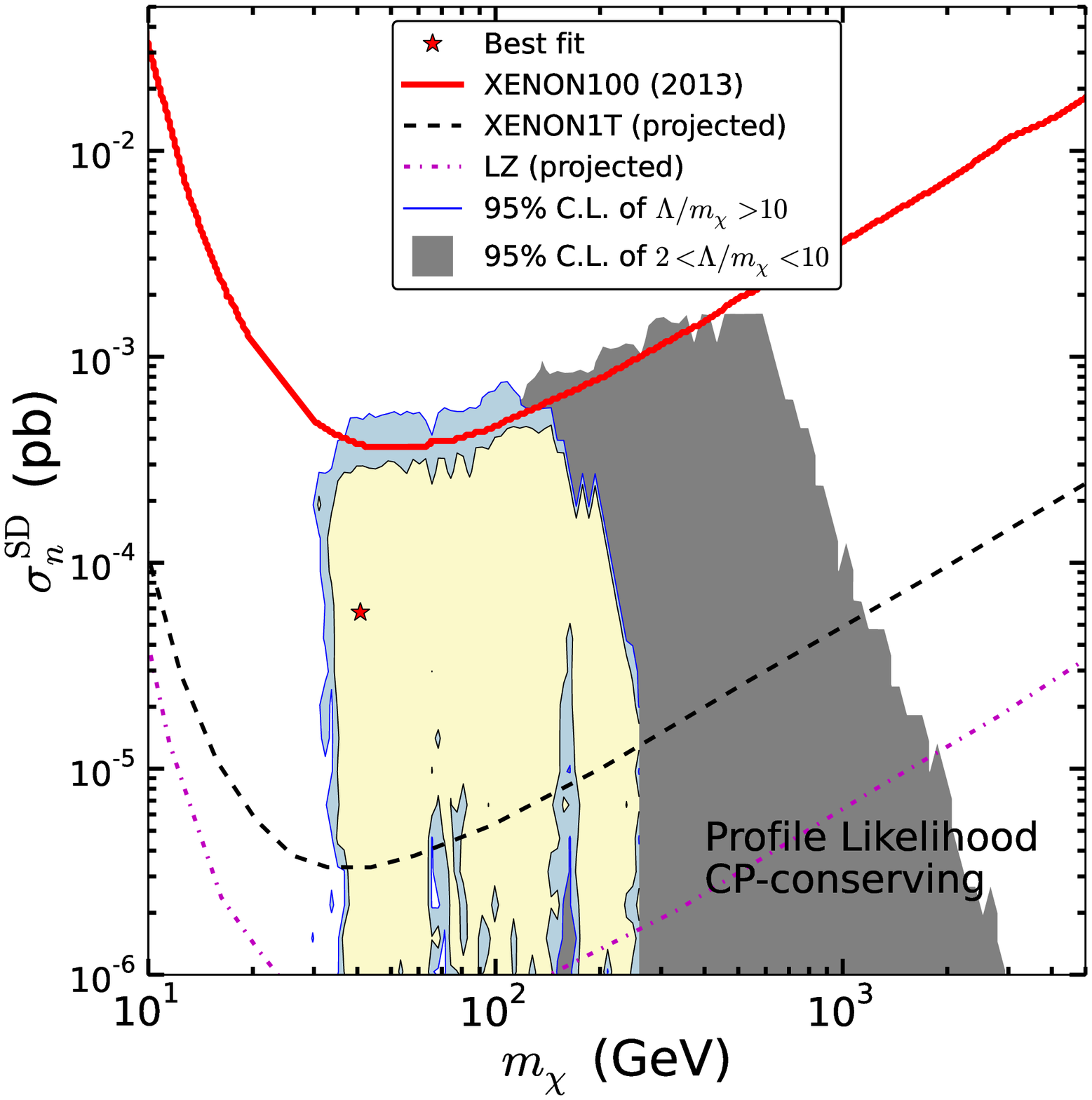}}
\caption{\small \sl SI (top-left) and SD (bottom panel) DM-nucleon scattering rates in the currently allowed EFT parameter space. The colour coding in the shaded regions is the same as in Fig.~\ref{CPC_mx_lam}. Current limits (LUX, XENON-100) as well as future projections (XENON-1T, LZ) are also shown. The cross-section for the process $e^+ e^- \rightarrow \chi \chi \gamma$  is shown in the top-right panel, along with the projection for the ILC experiment (see text for details).}
\label{CPC-obs}
\end{figure}

In Fig.~\ref{CPC-obs} (top-left) we show $\sigma_p^{\rm SI}$ as a function of $\mchi$ as found in the allowed EFT parameter space. The blue (yellow) shaded region is allowed at $95\%$ ($68\%$) C.L. after profiling over all the other parameters except $\mchi$ with the condition $\Lambda > 10 \mchi$. If the last condition is relaxed to $\Lambda > 2 \mchi$ the additional grey shaded region also survives at $95 \%$ C.L. There is a small difference between the $90 \%$ LUX constraint (red solid line) and the $95\%$ C.L. region obtained in our scan after including the LUX limits, primarily because the astrophysical parameters related to the DM local density and the DM velocity distribution are kept fixed in the LUX analysis, while they have been profiled out in our analysis including the errors as discussed in Sec.~\ref{sec3}. The future projections of the XENON-1T~\cite{Xenon1T} (black dashed line) experiment and the upgrade of the LUX experiment, LZ~\cite{LZ} (blue dashed line) are also shown in this figure, and they can be seen to cover upto $\sigma_p^{\rm SI} \sim 10^{-11}-10^{-12} $ pb . Finally, the yellow dashed line represents the ultimate sensitivity of the direct detection experiments, beyond which coherent neutrino-nucleon scattering will appear as an irreducible background~\cite{neutrino-nucleon,Snowmass}~\footnote{Including directional information may help in reducing the background from neutrino-nucleon scattering~\cite{Snowmass}.}. As we can see, in the Higgs-resonance region, $\sigma_p^{\rm SI} \sim 10^{-12}$ pb (as discussed in Sec.~\ref{CPC_all}), but for DM masses in which $\sv$ is dominated by other operators, lower values of $\sigma_p^{\rm SI} \lesssim 10^{-13}$ pb are allowed, which can even be below the floor of the neutrino-nucleon coherent scattering background.

Similarly, the SD scattering rates of DM with protons and neutrons are shown in the bottom left and right panels of Fig.~\ref{CPC-obs} respectively. The current strongest limit on $\sigma_n^{\rm SD}$ from XENON-100 is shown by the red solid line, while future projections from XENON-1T (black-dashed) and LZ (pink dot-dashed) are shown for both $\sigma_p^{\rm SD}$ and $\sigma_n^{\rm SD}$. Since the spin expectation value of the neutron group in XENON is much larger than that of the proton group, the constraints and projections for $\sigma_n^{\rm SD}$ are stronger. 

In Fig.~\ref{CPC-obs} (top-right) we show the cross-section for the process $e^+ e^- \rightarrow \chi \chi \gamma$ at the currently allowed parameter points of the EFT (at $68\%$ and $95\%$ C.L.). The cross-sections are shown after nominal selection cuts following the minimum requirements in the LEP mono-photon search ($E_\gamma > 6$ GeV and $3.8^\circ \lesssim \theta_\gamma \lesssim 176.2^\circ$), for the $e^+ e^-$ centre of mass energy of $1$ TeV. Since the EFT prediction of the cross-sections for $\Lambda < \sqrt{s}$ are not valid in general, in the pink-shaded region, which corresponds to $\Lambda< 1 \tev$ in this case, correct predictions can only be made in the UV-complete theory. However, as argued in Sec.~\ref{sec:LEP}, for lower DM masses, owing to the very large production cross-sections, some of these points might still be ruled out in the corresponding UV-completion models to the EFT, by a future $1 \tev$ $e^+e^-$ collider.  The red line represents the expected upper bound on the signal cross-section in the mono-photon$+\met$ process in an $e^+ e^-$ collider like the proposed ILC, with the luminosity fixed at $1 {~\rm ab}^{-1}$ (for other luminosities, the upper bound simply scales as $\sqrt{\mathcal{L}}$). We emphasize that the upper bound shown is just a simple estimate and only the selection efficiency of the above nominal cuts on the detected photon is included. After a particular detector design is adopted, the overall efficiencies will be modified, and the upper bound will be somewhat shifted. Therefore, Fig.~\ref{CPC-obs} gives us the order of magnitude expectation of $\sigma(e^+ e^- \rightarrow \chi \chi \gamma)$ in the allowed region of the EFT, and we see that cross-sections in the range of $2$ pb to $10^{-3}$ fb or lower can be possible. The upper bounds from future ILC measurements can be around $\mathcal{O}({\rm fb})$ and thus can cover a significant range of the possible cross-sections. It is clear that the resonance regions (Higgs or Z) may not yield high signal rates, as small couplings suffice to satisfy the relic abundance requirement.

\begin{figure}[htb!]
\begin{center}
\centerline{\epsfig{file=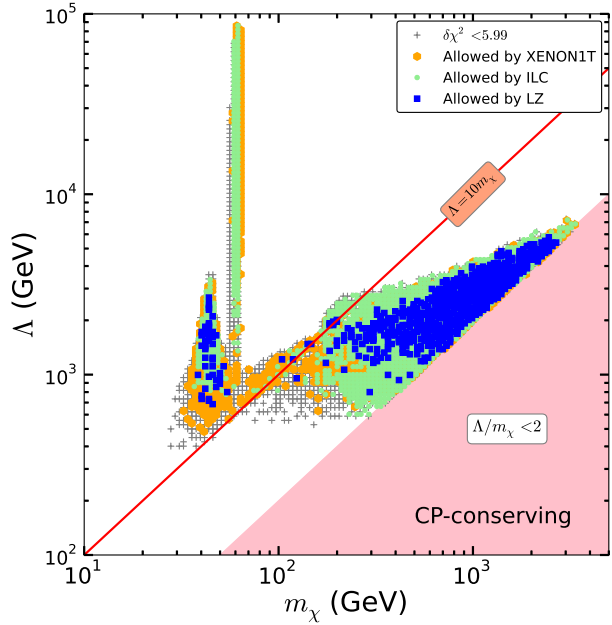
,width=9cm,height=9cm,angle=-0,keepaspectratio}}
\caption{\small \sl Parameter points in the $\mchi-\Lambda$ plane that survive after imposing the projected constraints from XENON-1T, ILC Higgs invisible width and mono-photon search, and the LZ experiments. The constraint from each experiment is imposed separately.}
\label{CPC_future}
\end{center}
\end{figure}

Finally, we show in Fig.~\ref{CPC_future} a combined view of the different possible future constraints, and the EFT parameter space in the $\mchi-\Lambda$ plane that survives at $95\%$ C.L. after applying each of them individually. The grey points marked as $\delta \chi^2 < 5.99$ represent the currently allowed parameter space. We first apply the projections of the XENON-1T experiment for SI and SD direct detection rates, and the yellow points are found to survive the cuts. The prospects of the ILC are considered next, whereby we include the expected constraints from a very precise measurement of the Higgs boson invisible branching ratio and from the mono-photon$+\met$ search. The sensitivity of the Higgs invisible branching ratio determination is expected  to reach $0.7\%$ level in the $ZH$ production mode after including the $Z\rightarrow q \bar{q}$ decay channel~\cite{Higgs-inv}. For the $\gamma+\met$ search in $e^+ e^-$ collisions, we have taken into account collisions with $\sqrt{s}=250,500 {~\rm and~} 1000$ GeV, in each case assuming an integrated luminosity of 
$1 {~\rm ab}^{-1}$. The reason for taking into account all the centre of mass energies is the fact that for a given $\sqrt{s}$, we can only reliably compute the production cross-sections in an EFT for $\Lambda>\sqrt{s}$. Therefore, only by taking into account the three different energies, we can cover an wide range of values of $\Lambda$. The green points remain after imposing the ILC projections, a large fraction of which are in the Higgs resonance region. However, this region can also be covered by the projected sensitivity of the LZ experiment, which is applied next, and the blue points survive the LZ projections of SI and SD scattering rates. 

For $\Lambda>10 \mchi$, the finally surviving points after considering all the future experiments, are concentrated in two regions. One of them is the Z-resonance region, and the other one is the region where the DM is heavier than 100 GeV. It is interesting to note that if the ILC can be operated at a centre of mass energy close to the Z-pole (the so-called Giga-Z option), then a significant part of the remaining points in the Z-resonance region can also be covered. For $2 \mchi \lesssim \Lambda \lesssim 10 \mchi$, a large number of points survive for $\mchi \gtrsim 200 \gev$, which cannot be covered by the future experiments considered in our study. 

The high-energy run of the LHC at 13 TeV, including its upgrade to a high-luminosity phase, may be able to cover part of the last mentioned region above using the mono-jet$+\met$ channel. However, usually we require strong kinematical cuts at $13\tev$ LHC in order to suppress the very large background coming from the $Z(\rightarrow \nu \bar{\nu})+{\rm jets}$ process, which essentially translates to a requirement of high sub-process centre of mass energies in the events (of the order of 1 TeV or higher). Since $\Lambda$ in this region also lies in the ballpark of a TeV, it is difficult to estimate the future LHC reach in this part of the parameter space within the EFT. A proper estimate can only be made by using suitable simplified models representing the effective operators (with s- or t-channel mediators between the DM and the SM sector). Such a study is, however, beyond the scope of the present article.

\section{Results: CP-violating scenario}
\label{sec5}
\begin{figure}[htb!]
\begin{center}
\centerline{\epsfig{file=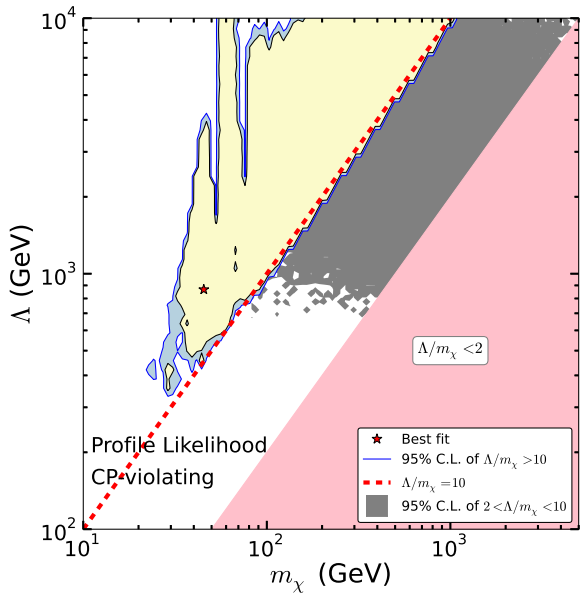
,width=8.5cm,height=8.5cm,angle=-0,keepaspectratio}}
\caption{\small \sl Same as Fig.~\ref{CPC_mx_lam}, in the CP-violating scenario, which includes the additional Higgs-portal operator in Eq.~\ref{dim5cpv}.}
\label{CPV_mx_lam}
\end{center}
\end{figure}
We now summarize the results in the CP-violating scenario, where one additional operator, $\mathcal{L}_5^{\rm {CPV}}$ in Eq.~\ref{dim5cpv} is also included. This additional operator helps enhancing $\sv$ considerably, and therefore opens up a much larger parameter space along the $\mchi$ direction. The constraints from SI and SD scattering rates remain the same as in the CP-conserving scenario, since the DM bilinear $\overline{\chi} \gamma_5 \chi$ vanishes in the non-relativistic limit. We show the currently allowed parameter space in the $\mchi-\Lambda$ plane in Fig.~\ref{CPV_mx_lam}, with the same colour coding as in Fig.~\ref{CPC_mx_lam}. We have already seen the impact of $\mathcal{L}_5^{\rm {CPV}}$ in determining $\Omega_\chi h^2$ in Sec.~\ref{relic-role}, where we found that in the entire mass range of $10 \gev \lesssim \mchi \lesssim 1 \tev$ the required $\sv$ can be obtained for $\Lambda>10 \mchi$. However, in the low DM mass region, $\mchi \lesssim 30$ GeV, the Higgs invisible width via $\mathcal{L}_5^{\rm {CPV}}$ is larger than that allowed by current constraints, and therefore, this region remains ruled out in the CP-violating scenario as well. But beyond the threshold of $\mchi>M_h/2$, no other current constraint on this operator is strong enough, and therefore a large region of additional parameter space with $70 \gev \lesssim \mchi \lesssim 1 \tev$ opens up. The gamma ray constraints from dwarf spheroidal observations by Fermi-LAT do play some role in constraining this CP-violating coupling due to the lifting of the p-wave suppression. We find that in this additional region, not only does the coupling $g_{\rm PS}$ determine the relic density, a small, but LUX-allowed value of $g_{\rm S}$ comes into play as well. 

\begin{figure}[htb!]
\centering
\centerline{\includegraphics[width=8.5cm,height=8.5cm,keepaspectratio]{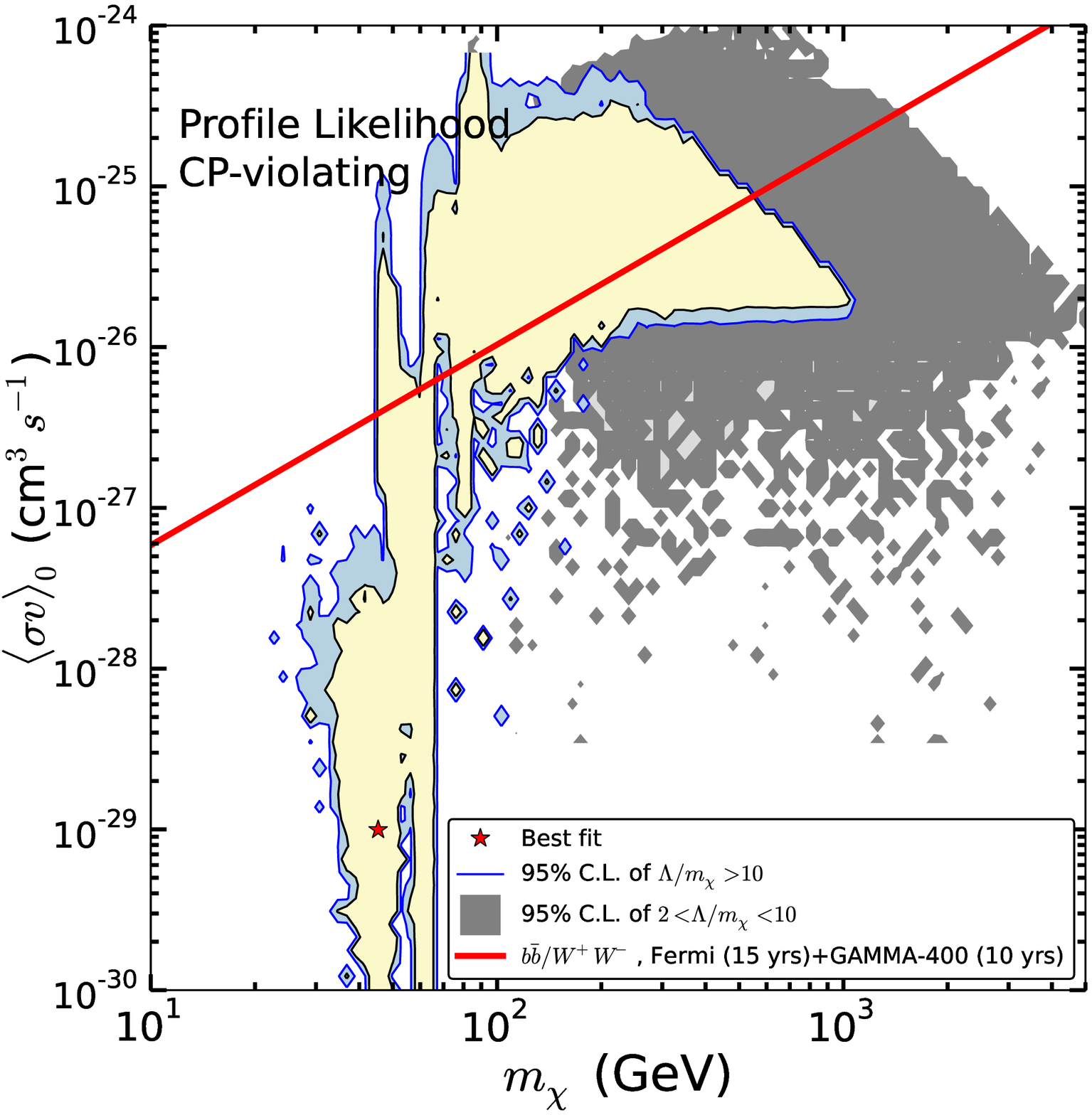}\
\includegraphics[width=8.5cm,height=8.5cm,keepaspectratio]{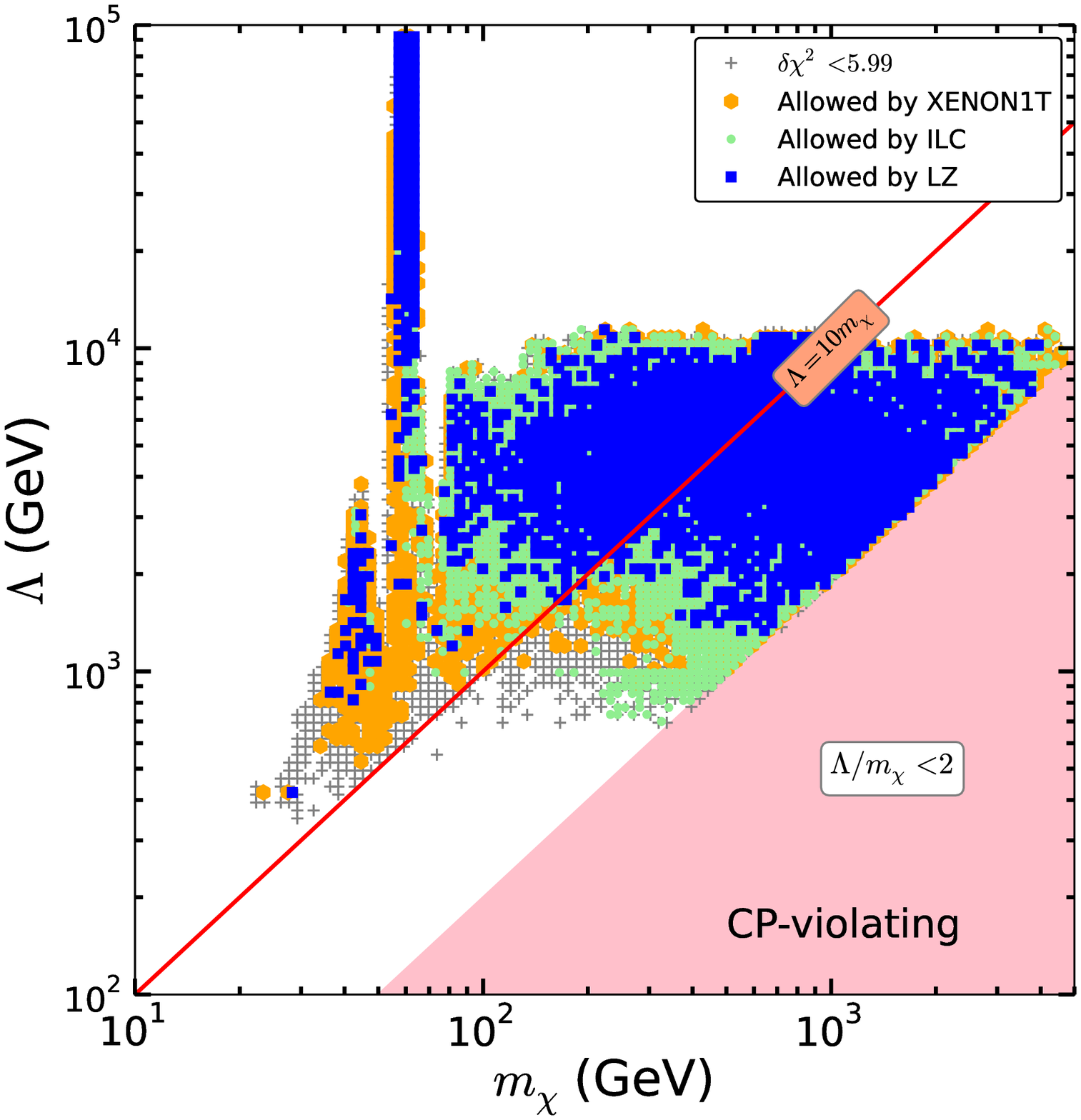}}
\caption{\small \sl Annihilation rate in the present universe,  $\langle \sigma v \rangle_0$ (left panel) and parameter points in the $\mchi-\Lambda$ plane that survive after imposing the projected constraints from XENON-1T, ILC Higgs invisible width and mono-photon search, and the LZ experiments (right panel), in the CP-violating scenario.}
\label{fig:CPV-obs}
\end{figure}
Since the predictions for the SI and SD scattering rates are very similar in the CP-violating scenario, we do not show them separately. For the annihilation rate in the present universe, due to the lifting of the helicity suppression in the s-wave annihilation amplitudes to SM fermion pairs, considerably higher values of $\langle \sigma v \rangle_0$ can be obtained. We show in Fig.~\ref{fig:CPV-obs} (left panel) the values of $\langle \sigma v \rangle_0$ as a function of $\mchi$ in the EFT parameter space. The red solid line represents the future sensitivity of gamma ray observations from dwarf spheroidal galaxies using the Fermi-LAT (15 years of data taking) and the proposed GAMMA-400~\cite{gamma400} (10 years of data taking) experiments, in the $\chi \chi \rightarrow W^+ W^-$ channel, as estimated in Ref.~\cite{Shigeki-dwarf}. For $\mchi<M_W$, the $\chi \chi \rightarrow b \bar{b}$ channel is relevant, which gives rise to similar constraints as well. This sensitivity line is only indicative of the future reach, and will vary depending upon the specific annihilation channel(s) relevant in different parameter regions of the EFT. 

In Fig.~\ref{fig:CPV-obs} (right panel) we show the surviving parameter points in the CP-violating scenario, in the $\mchi-\Lambda$ plane. The various constraints imposed in this figure are discussed in the context of the CP-conserving scenario (Sec.~\ref{sec:CPC-fut}). In addition to the Z-resonance region and a much larger bulk region with $\mchi>100 \gev$, an additional set of points in the Higgs-resonance region also survives all the future experiments considered, since the CP-violating operator is not constrained by SI scattering rates, which, for the CP-conserving case, could completely cover the Higgs-resonance part. If indeed a CP-violating operator exists in the DM sector, it is then imperative to find out other possible experiments which can probe such a coupling.

\section{Summary}
To summarize, we studied a standard model singlet Majorana fermion DM candidate in the framework of an effective field theory, by taking into account the presence of a complete basis of gauge invariant operators of dimensions 5 and 6 at the same time. The profile likelihood method was used to determine the currently allowed region of parameter space at $95\%$ C.L., and the span of different DM related observables as a function of its mass. Considering an weakly coupled ultra-violet completion for the EFT, we find that imposing the condition $\Lambda > 10\mchi$ makes the EFT an excellent approximation to possible UV complete models. For completeness, we have also computed the confidence intervals in the region $ 2\mchi <\Lambda < 10\mchi$. Including the $\mathcal{O}(1)$ coefficients in front of each operator (which can be either of positive or negative sign), the likelihood analysis is performed over the 8-dimensional parameter space in a CP-conserving scenario, and a 9-dimensional one once the singlet CP-violating scalar interaction with the Higgs boson is also switched on. The astrophysical and nuclear physics parameters involved in computing the DM observables are treated as nuisance parameters, and are profiled out. Constraints from various experiments are taken into account: relic abundance (as an upper bound) from {\it Planck}, SI scattering rates from LUX, SD ones from XENON100 and IceCube, gamma-ray constraints from Fermi-LAT, Z-invisible width and mono-photon limits from LEP and Higgs invisible branching ratio and mono-jet limits from the 8 TeV LHC. For future projections, we consider the capabilities of XENON1T, LZ, ILC, Fermi-LAT and GAMMA-400. This study, to our knowledge, constitutes the first comprehensive analysis of a singlet Majorana fermion DM in an EFT.

The primary results in the CP-conserving case can be summed up as follows. A DM of mass $\mchi<30 \gev$ looks disfavoured when all constraints are put together. Primarily because of the upper bound on the thermal component of the relic density, we also have an upper bound on the DM mass, $\mchi \lesssim 300 \gev$ is allowed if $\Lambda>10 \mchi$. This is relaxed to $\mchi \lesssim 3 \tev$ if $\Lambda>2 \mchi$. If the DM mass is close to half the Z-boson mass or half the Higgs boson mass, the allowed values of $\Lambda$ are very high, especially in the latter case, where it can go upto $\mathcal{O}(100 \tev)$. For $\mchi>70 \gev$, only values of $\Lambda$ very close to $10\mchi$ are allowed, and a little more area opens up when the DM mass crosses the top-quark threshold.

Among the presently allowed parameter region, surprisingly enough, it seems viable that the Higgs resonance region will be completely covered by the combined efforts of future ton-scale direct detection experiments like XENON1T and LUX, as well as the measurement of the Higgs invisible branching ratio at the proposed International Linear Collider. The main difference between the Higgs and Z-resonance regions is that the former leads to SI scattering rates, while the latter to SD ones. Therefore, the direct detection experiments will probably not be able to cover whole of the Z-resonance region with their currently projected sensitivities. Here also, the ILC can play a major role, when we use the mono-photon search channel for DM pair-production. For DM heavier than the top mass, where the four-fermion operators play a major role, there is a large set of allowed points if $2 \mchi < \Lambda < 10\mchi$. If the couplings to quarks are flavour universal, the high energy and high luminosity LHC runs should have an impact here. However, it is difficult to judge it within an EFT, especially if only the weak condition $\Lambda>2\mchi$ holds. We leave a detailed study of such a region within realistic simplified models with s- and t-channel mediator exchange  for a future work.

If the CP-violating operator is present, the allowed parameter region is larger, since DM pair annihilation to light quarks are no longer p-wave suppressed and contribute to $\Omega_\chi h^2 {~\rm (Thermal)}$. DM masses of upto about 1 TeV now become viable even with $\Lambda>10\mchi$. The same reason also gives us a hope of testing at least part of the parameter space using gamma-ray observations from DM annihilation to quarks and leptons. We presented some estimates of this indirect detection capability, and further detailed studies are necessary here as well. It also seems necessary to pursue new avenues to probe the CP-violating coupling itself, since it is not testable using low-energy scattering experiments with nuclei.

\section*{Appendix A: The direct detection likelihood}
\label{apA}
As discussed in Sec.~\ref{DD}, the LUX collaboration obtained the bounds on $\sigma_p^{\rm SI}$  by fixing the astrophysical inputs to specific values and by taking the velocity distribution of DM to be  a truncated Maxwellian~\cite{LUX}. Since we use the DM phases-space density computed in Ref.~\cite{Ullio} along with its $2\sigma$ error bars, we first rescale the LUX bounds accordingly. For calculating the time-averaged number of events we adopted the Helm form factor \cite{formf} for $\sigsip$ as used by LUX, together with the energy resolution function and efficiency factors for the LUX detector~\cite{LUX}. In the following, the rescaled values of the LUX $90\%$ C.L. bounds as a function of the DM mass are denoted by $\sigma^{\rm{SI}}_{p,90\%}$. A similar approach is applied to rescale the XENON100~\cite{XENON} bounds on $\sigsdn$ as well, with the corresponding form factors relevant for SD scattering~\cite{Engel:1991wq}. 

In constructing the likelihood, the nuclear physics inputs discussed in Sec.~\ref{DD}, namely $f_{Ts}$ and $\Sigma_{\pi N}$ for $\sigsip$, and 
$\Delta q^{n,p}$ for $\sigsdn$, are treated as nuisance parameters. Together with the rescaled $90\%$ C.L. limit from LUX for $\sigsip$ and from
XENON100 for $\sigsdn$ we construct the likelihood functions as follows~\cite{Cheung:2012xb,Fowlie:2013oua}:
\begin{eqnarray}
\label{eq:xenonlike}
 \mathcal{L}_{\rm LUX} &\propto& \exp [ -\frac{1}{2}
 \frac{(\sigsip-0.0)^2}
 {(\sigma^{\rm{SI}}_{p,90\%}/1.64)^2 +\delta_{sys.}^2} ] 
 \; \label{eq:Llux},\\
 \mathcal{L}_{\rm X100} &\propto& \frac{e^{-(s + b')}\left(s + b' \right)^{o}}{o !} 
\exp [ -\frac{(b'-b)^2}{2\sigma_b^2} ] 
 \; \label{eq:LX100}. 
\end{eqnarray}
In Eq. (\ref{eq:Llux}), we assume zero signal events and hence set the central value of $\sigsip$ to zero. The $1\sigma$ error is taken as $\sigma^{\rm{SI}}_{p,90\%}/1.64$. The blue shaded band in Fig.~5 of the LUX published results~\cite{LUX} gives us an estimate of the systematic uncertainty, $\delta_{sys.}$. The theoretical error from nuclear matrix elements enters the computation of $\sigsip$. For the XENON100 likelihood in Eq. (\ref{eq:LX100}), the expected number of background events is taken as $b=1.0\pm 0.2$, and the number of observed events is $o=2$ as reported in the XENON100 (2012) data \cite{XENON}. 

\section*{Appendix B: Monojet$+ \met$ search at the LHC}
\label{apB}
In the CMS search for monojet$+\met$ with the $8$ TeV LHC data, the primary selection criterion used are~\cite{CMS}:
\begin{enumerate}
\item At least one jet $j_1$ within a pseudo rapidity of $|\eta|<2.4$, and transverse momentum $p_T>110$ GeV. 

\item The optimized value of $\met$ as determined by the CMS collaboration: $\met > 400$ GeV.

\item A second hadronic jet $j_2$ is allowed if its azimuthal separation from the highest $p_T$ jet satisfies: $\Delta \phi(j_1 j_2) < 2.5$. Since in the major background process of QCD dijets, the jets are produced back to back in the transverse plane, this angular requirement, together with the demand for a large $\met$, helps in reducing this background considerably.

\item An event with a third additional jet with $p_T>30$ GeV and $|\eta|<4.5$ is discarded, as are events with isolated charged leptons.
\end{enumerate}
\begin{figure}[htb!]
\begin{center}
\centerline{\epsfig{file=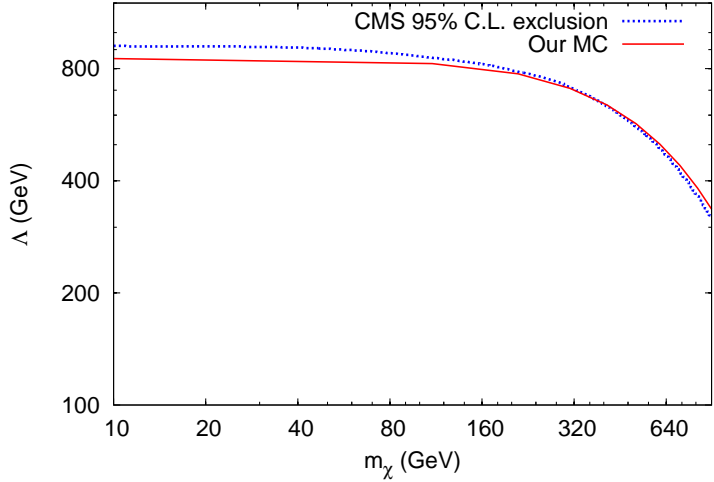
,width=10cm,height=8cm,angle=-0,keepaspectratio}}

\caption{\small \sl Comparison of our MC simulation and CMS results for the $95\%$ C.L. upper bound in the $\mchi-\Lambda$ plane using the monojet$+\met$ channel, for an axial-vector DM-quark interaction (see text for details).}
\label{CMS-compare}
\end{center}
\end{figure}
As discussed in Sec.~\ref{LHC}, we performed the MC simulation for the above search for each point in our EFT parameter space. In order to validate our MC, we first compare the bounds obtained for a Dirac fermion DM with an axial vector interaction with quarks, $(\overline{\chi} \gamma_{\mu} \gamma^5 \chi) (\overline{q} \gamma^{\mu} \gamma^5 q)$, with the CMS results, and find agreement to better than $5\%$. The MC simulation framework used by us follows the {\tt FeynRules-MadGraph5-Pythia6}~\cite{Pythia}{\tt -Delphes2}~\cite{Delphes} chain, with the jets reconstructed using the {\tt anti-$k_T$} algorithm~\cite{antikt} as implemented in {\tt FastJet2}~\cite{Fastjet}, with a cone size of $R=0.4$. We used the {\tt CTEQ6L1}~\cite{Cteq} parton distribution functions with the factorization and renormalization scales set at the default dynamical scale choice of {\tt MadGraph5}. The comparison of our simulation with the CMS results is shown in Fig.~\ref{CMS-compare}.

The likelihood function for the LHC monojet search can be expressed as follows
\begin{equation}
\mathcal{L}(N_{\rm obs}|b+s) \propto \max_{b^\prime=\{0,\infty\}}\frac{e^{-(s + b')}\left(s + b' \right)^{N_{\rm obs}}}{N_{\rm obs} !} \exp [ -\frac{(b'-b)^2}{2\sigma_b^2}]~, 
\label{LHC-like}
\end{equation}
where, $b$ is the expected number of background events, $s$ is the expected number of signal events for a given parameter point and $N_{\rm obs}$ represents the number of events observed by CMS after employing the kinematic selection criterion described above. The systematic uncertainty is taken into account by convoluting the Poission likelihood function with a Gaussian with mean $b$ (which is treated as a nuisance parameter and profiled out) and variance $\sigma_b$.

\section*{Acknowledgments}
We thank Norimi Yokozaki, Myeonghun Park, Qiang Yuan and Xiaoyuan Huang for useful discussions. This work is supported by the Grant-in-Aid for Scientific Research from the Ministry of Education, Science, Sports, and Culture (MEXT), Japan (No. 26287039, 22244031 and 26104009 for S. Matsumoto), and by the World Premier International Research Center Initiative (WPI Initiative), MEXT, Japan.


\end{document}